\documentclass[fleqn,usenatbib]{mnras} %original
\usepackage{newtxtext,newtxmath}

\usepackage[T1]{fontenc}
\usepackage{ae,aecompl}

\usepackage[dvipdfmx]{graphicx}
\usepackage{amsmath}	% Advanced maths commands
\usepackage{amssymb}	% Extra maths symbols
\usepackage{bm}
\usepackage{setspace}
\usepackage{lscape}

\newcommand{\gturb}{\gamma_{\rm turb}}
\newcommand{\beff}{\beta_{\rm eff}}
\newcommand{\msun}{\thinspace M_\odot}
\newcommand{\gcm}{~{\rm g~cm}^{-3}}

\newcommand{\mach}{\cal{M}}

\newcommand{\FiducialModel}{D4 ($\alpha = 0.5$, $\gturb = 0.1$) }
\newcommand{\WarpModel}{B5 ($\alpha = 0.3$, $\gturb = 0.3$) }
\newcommand{\Js}{\bm{J}_{\rm s}}
\newcommand{\Jd}{\bm{J}_{\rm d}}
\newcommand{\Je}{\bm{J}_{\rm e}}
\newcommand{\Jshell}{\bm{J}_{{\rm shell}}}
\newcommand{\psd}{\psi_{\rm sd}}
\newcommand{\pse}{\psi_{\rm se}}
\newcommand{\pde}{\psi_{\rm de}}
\newcommand{\pshell}{\psi_{\rm shell}}
\title[Star-disk alignment from SPH simulation] {Star-disk alignment
  in the protoplanetary disks: SPH simulation of the collapse of
  turbulent molecular cloud cores}

\author[D. Takaishi et al.]{
  Daisuke Takaishi,$^{1}$\thanks{E-mail: k3790238@kadai.jp (D. Takaishi)}
  Yusuke Tsukamoto,$^{1}$
  and Yasushi Suto$^{2,3}$ \\
  $^1$Graduate School of Science and Engineering,
  Kagoshima University, Kagoshima 890-0065, Japan\\
  $^2$Department of Physics, The University of Tokyo, Tokyo 113-0033, Japan\\
  $^3$Research Center for the Early Universe,
  School of Science, The University of Tokyo, Tokyo 113-0033, Japan\\
}

\date{Accepted 2020 January 14. Received 2020 January 8; in original
  form 2019 August 10}

\pubyear{2020}

\begin{document}
\label{firstpage}
\pagerange{\pageref{firstpage}--\pageref{lastpage}}
\maketitle

\begin{abstract}
  \label{abst}
  We perform a series of three-dimensional smoothed particle
  hydrodynamics (SPH) simulations to study the evolution of the angle
  between the protostellar spin and the protoplanetary disk rotation
  axes (the star-disk angle $\psd$) in turbulent molecular cloud
  cores.  While $\psd$ at the protostar formation epoch exhibits broad
  distribution up to $\sim 130^{\circ}$, $\psd$ decreases ($\lesssim
  20^{\circ}$) in a timescale of $\sim 10^{4}$ yr.  This timescale of
  the star-disk alignment, $t_{\rm alignment}$, corresponds basically
  to the mass doubling time of the central protostar, in which the
  protostar forgets its initial spin direction due to the mass
  accretion from the disk.  Values of $\psd$ both at $t=10^2$ yr and
  $t=10^5$ yr after the protostar formation are independent of the
  ratios of thermal and turbulent energies to gravitational energy of
  the initial cloud cores: $\alpha=E_{\rm thermal}/|E_{\rm gravity}|$
  and $\gturb=E_{\rm turbulence}/|E_{\rm gravity}|$.  We also find
  that a warped disk is possibly formed by the turbulent accretion
  flow from the circumstellar envelope.
\end{abstract}


%%%%%%%%%%%%%%%%%%%%%%%%%%%%%%%%%%%%%%%%%%%%%%%%%%

%%%%%%%%%%%%%%%%% BODY OF PAPER %%%%%%%%%%%%%%%%%%

\section{Introduction}
\label{sec:intro}
%%%%%%%%%%%%%%%%%%%%%%%%%%%%%%%%%%%%%%%%%%%%%%%%%%

Observed exoplanetary systems have exhibited unexpectedly broad
diversities \citep{2015ARA&A..53..409W}.  One of the intriguing
discoveries is the fact that approximately 20\% of hot Jupiter have
orbital planes misaligned relative to the spin axis of their host
stars.  For instance, \citet{2019AJ....157..137K} shows that 28 out of
124 transiting close-in gas-giant planets have the projected
spin-orbit angle $\lambda$ exceeding $30^{\circ}$ via the Rossiter
McLaughlin (RM) effect \citep{1924ApJ....60...15R,
  1924ApJ....60...22M, 2000A&A...359L..13Q, 2005ApJ...622.1118O,
  2005ApJ...631.1215W, 2011ApJ...742...69H, 2012ApJ...757...18A,
  2018haex.bookE...2T}.

The origin of the large spin-orbit angle remains unclear.  One of the
promising mechanism is the dynamical evolution of the orbital plane by
planet-planet and star-planet interactions.  Because the RM effect has
been preferentially observed for short-period and giant planets, the
violent dynamical evolution such as the planetary migration
\citep[e.g.,][]{1996Natur.380..606L, 2005A&A...434..343A},
planet-planet scattering \citep[e.g.,][]{1996Sci...274..954R,
  2008ApJ...678..498N, 2011ApJ...742...72N, 2012ApJ...751..119B}, and
strong perturbation due to distant outer objects
\citep[e.g.,][]{1962AJ.....67..591K, 1962P&SS....9..719L,
  2007ApJ...669.1298F, 2012Natur.491..418B, 2014ApJ...784...66X,
  2016MNRAS.456.3671A, 2016ApJ...820...55X} possibly explains the
large spin-orbit angle.  According to these mechanisms,
multi-planetary transiting systems which have almost co-planar orbital
planes may not have the significant star-planet misalignment because
the violent dynamical evolution also causes the misalignment between
the orbital planes of planets.  Consistent with this expectation,
Kepler-89 (with four transiting planets) and Kepler-25 (with two
transiting and one non-transiting planets) are suggested to have
$\lambda \sim 0$ from the RM observations by
\citet{2012ApJ...759L..36H} and \citet{2013ApJ...771...11A},
respectively.

On the other hand, however, there is a transiting multi-planetary
system, Kepler-56, which has a significant oblique stellar spin
although the planets in the system have almost co-planer orbits;
\citet{2013Sci...342..331H} showed that its stellar inclination angle
$i_{s}$ is $\sim 45^{\circ}$ from the asteroseismic analysis.  While
it could be explained by some kind of perturbation that changes the
two planetary orbits in a coherent fashion, it seems natural to
interpret it in terms of a primordial origin.  For instance, the
stellar spin axis may be significantly misaligned with the
protoplanetary disk rotation axis.

This possibility has been investigated in several previous studies.
\citet{2010MNRAS.401.1505B} approached the problem using smoothed
particle hydrodynamics (SPH) combined with the sink particle
technique. Specifically they followed evolution of a relative angle
between the stellar spin and the protoplanetary disk rotation axes
(hereafter, the star-disk angle $\psd$) in a star cluster that forms
from a supersonic turbulent molecular cloud with its mass, size, and
Mach number being $50\msun$, 0.375 pc (=77400 au), and ${\cal M}=6.4$,
respectively.  Although the star-disk angle $\psd$ can be misaligned
via the stellar close-encounter in a multiple star-forming region,
they pointed out that such events are rare and the orientations of the
disk and star tends to be aligned in most cases. Furthermore, they
suggested that the reliable prediction of the star-disk angle
distribution is not easy because the process occurs in an inherently
chaotic environment of the cluster forming region.

More recently, \citet{2015MNRAS.450.3306F} examined the evolution of
the star-disk angle $\psd$ in a massive molecular cloud with
supersonic turbulence, which has the mass, size, and Mach number of
$150\msun$, 0.397 pc (=81920 au), and ${\cal M}=7.5$,
respectively. They performed the hydrodynamic and magnetohydrodynamic
simulations with the grid-based adaptive mesh refinement (AMR), and
indicated that the large star-disk angles around $40^\circ$ are more
common. This results are consistent with the observed spin-orbit angle
distribution of hot Jupiters.  They confirmed that the gravitational
torque from the protoplanetary disk to the stellar quadrupole does not
wipe out the misalignment as long as the spin rate of the protostar is
significantly slower than the breakup rotation rate.

Both papers mentioned above focused on massive compact molecular
clouds with supersonic turbulence, which correspond to star-cluster
forming regions such as the Orion Nebula Cluster
\citep[e.g.,][]{1997AJ....113.1733H} and infrared-dark clouds
\citep{2012ApJ...754....5B}.  In nearby star-forming regions such as
the Taurus molecular cloud, however, a relatively compact and isolated
protostar forms from a low mass molecular cloud core. For instance,
the pre-stellar core L1544 is estimated to have mass of $\sim
1.3\msun$, number density of $\sim 4.9 \times 10^5~\rm{cm^{-3}}$, size
of $\sim 0.021$ pc, and velocity dispersion of $\sim
0.28~\rm{km~s^{-1}}$ \citep[e.g.,][]{1998ApJ...504..900T,
  1999ApJ...513L..61W, 2004ApJ...600..279C, 2007prpl.conf...33W}.
Therefore, the significant difference of the environment between
star-cluster forming regions and the nearby star-forming region may
affect the distribution of the star-disk angle.

In this paper, we focus on isolated turbulent molecular cloud cores
with typical sizes of 0.01-0.1 pc ($\sim1000-10000$ au) that have not
yet been explored in the above studies.  Because several observations
suggest that the molecular cloud cores have weak turbulence of ${\cal
  M}<1$ \citep[e.g.,][]{1996A&A...314..625A, 2007prpl.conf...33W}, we
consider sub- to trans-sonic turbulent molecular cloud cores and
examine the evolution of $\psd$, the angle between the protostar spin
and the protoplanetary disk rotation axes.

We neglect the magnetic field and start our simulations from a
spherically symmetric isothermal cloud core with the turbulent motion
following the power spectrum of $P_{v}(k) \propto k^{-4}$.  We perform
26 different simulations by varying their initial thermal and
turbulent energies.  We use the sink particle technique to represent
protostars, and examine the star and disk evolution for approximately
$10^{5}$ yr after the protostar formation.

The structure of this paper is as follows.  Section
\ref{sec:Method_and_IC} describes our numerical method and initial
conditions for the SPH simulation.  Section \ref{sec:D4-result}
discusses the results of our fiducial model in detail, with particular
attention to the evolution of the relative angles of orientations of
the protostar, protoplanetary disk, and the surrounding envelope
component.  Statistical analysis for 20 models having a single
protostar is presented in Section \ref{sec:statistical-result}.
Further implications of the present simulation are discussed in
Section \ref{sec:other_implications}, and finally Section
\ref{sec:conclusion} is devoted to the conclusion of this paper.

%%%%%%%%%%%%%%%%%%%%%%%%%%%%%%%%%%%%%%%%%%%%%%%%%%

%%% 2. numerical simulation
\section{Numerical Method and Initial Conditions of the Simulations}
\label{sec:Method_and_IC}
%%%%%%%%%%%%%%%%%%%%%%%%%%%%%%%%%%%%%%%%%%%%%%%%%%%%%%%%%%%%%

%%% 2.1. method
\subsection{Numerical Method}
\label{subsec:method}
%%%%%%%%%%%%%%%%%%%%%%%%%%%%%%%%%%%%%%%%%%%%%%%%%%%%%%%%%%%%%

We solve equations of hydrodynamics including self-gravity with the
smoothed particle hydrodynamics (SPH) method
\citep{1977AJ.....82.1013L, 1977MNRAS.181..375G, 1985A&A...149..135M},
%%%%%%%%%%%%%%%%%%%%%%%%%%%%%%
\begin{align}
  %\frac{D\rho}{Dt}   &= - \rho \nabla \cdot \bm{v},              \label{eq:continuity} \\
  \frac{D\bm{v}}{Dt} &= - \frac{1}{\rho} \nabla P - \nabla \phi, \label{eq:of_motion}  \\
  \nabla^{2} \phi    &=   4 \pi G \rho,                          \label{eq:poisson}
\end{align}
%%%%%%%%%%%%%%%%%%%%%%%%%%%%%%
where $\rho$ is the gas density, $\bm{v}$ is the gas velocity, $P$ is
the gas pressure, $\phi$ is the gravitational potential and $G$ is the
gravitational constant.  The SPH code that we use here has been
applied for a variety of problems \citep[e.g.,][]{2011MNRAS.416..591T,
  2013MNRAS.428.1321T, 2013MNRAS.436.1667T, 2015MNRAS.446.1175T,
  2015MNRAS.452..278T, 2015ApJ...810L..26T, 2016ApJ...833..105Y,
  2017PASJ...69...95T, 2018ApJ...868...22T}.

We adopt the barotropic equation of state,
%%%%%%%%%%%%%%%%%%%%%%%%%%%%%%
\begin{align}
  \label{eq:eos}
  P = c^2_{\rm{s,0}}\rho \left[ 1+\left(\frac{\rho}{\rho_{\rm c}}\right)^{2/5} \right],
\end{align}
%%%%%%%%%%%%%%%%%%%%%%%%%%%%%%
where $c_{\rm{s,0}} = 1.9 \times 10^{4} \rm{~cm~s^{-1}}$ is the sound
velocity at the temperature of $10~{\rm K}$ and $\rho_{\rm c} =
4\times 10^{-14} {\rm \gcm}$ is the critical density at which the
thermal evolution changes from the isothermal to adiabatic.  This
empirical equation of state is adopted in previous disk formation
simulations neglecting the radiation transfer
\citep[e.g.,][]{2007ApJ...670.1198M, 2010ApJ...724.1006M,
  2013MNRAS.428.1321T}.  The molecular cloud core is assumed to have
an initial temperature of $T=10~{\rm K}$ at which the cosmic-ray
heating is balanced with the cooling of the molecular line emissions
and dust continuum emissions \citep[e.g.,][]{2007ARA&A..45..565M,
  2017iace.book.....Y}.

The main purpose of the present study is to examine the angle between
the stellar spin and the disk rotation axes $\psd$.  However, it is
impossible to numerically resolve the central protostar.  Therefore we
adopt the sink particle technique \citep{1995MNRAS.277..362B} and
regard the mass and spin direction of the sink particle as those of
the protostar.  We create a sink particle when the density of SPH
particle reaches the threshold value $\rho_{\rm
  sink}=4\times10^{-8}\gcm$, which corresponds to the density when the
second collapse begins \citep[e.g.,][]{2000ApJ...531..350M,
  2012PTEP.2012aA307I}.  The sink particle interacts with SPH
particles through gravity.  We set the accretion radius of the sink
particle as $r_{\rm acc}=1$ au, and all the SPH particles within the
accretion radius are removed, and their mass, linear momentum, and
angular momentum with respect to the sink particle are added to the
sink particle.  The accretion radius of 1 au can reasonably resolve
the formation and early evolution of the protoplanetary disk
\citep{2014MNRAS.438.2278M}.  Note that this accretion radius is much
smaller than 5 au adopted by \citet{2010MNRAS.401.1505B}.

We simply add the accreted mass and angular momentum to the sink
particle.  While this procedure conserves the angular momentum within
the radius of 1 au represented by the sink particle, it should {\it
  not} be identified with the spin angular momentum of the protostar
itself because it exceeds the breakup value.
\citet{2010MNRAS.401.1505B} and \citet{2015MNRAS.450.3306F} proposed
different schemes of estimating the stellar spin on the basis of the
imposed sub-grid physics. As described in the next subsection, we
implemented the procedure by \citet{2015MNRAS.450.3306F}, and
re-simulated one of the model. We confirmed that their scheme
significantly suppresses the amplitude of the spin, but that its
direction is almost unchanged. Therefore, we decided to use the total
angular momentum vector within the radius of 1 au from the sink
particle as a good proxy for the {\it direction} of the central
stellar spin.

%%%%%%%%%%%%%%%%%%%%%%%%%%%%%%%%%%%%%%%%%%%%%%%%%%%%%%%%%%%%%

%%% 2.2. initial conditions / set up
\subsection{Initial Conditions}
\label{subsec:IC}
%%%%%%%%%%%%%%%%%%%%%%%%%%%%%%%%%%%%%%%%%%%%%%%%%%%%%%%%%%%%%

The hydrodynamic simulations with both magnetic field and turbulence
are computationally very demanding, and it is not easy to perform the
parameter study as attempted below. Thus we decide to ignore the
magnetic field in the present simulation, and focus on the effect of
the turbulence on the spin-orbit architecture of the protoplanetary
disks.  The simulation follows approximately $10^{5}$ yr after the
protostar formation.  We plan to incorporate the magnetic field in the
subsequent work.

For the initial condition, we adopt spherically symmetric and
isothermal cloud cores with the turbulent velocity field which obeys
the velocity power spectrum of $P_{v}(k) \propto k^{-4}$
\citep{1993ApJ...406..528G, 1998ApJ...504..207B, 2000ApJ...543..822B}.
The total mass of the cloud core is fixed to be $1M_{\odot}$.  The
number of SPH particles is $N_{\rm p}\sim 10^6$ and the mass of SPH
particles is set to be $m_{\rm sph} = 1M_{\odot}/N_{\rm p} = 10^{-6}
M_{\odot}$.  \citet{1997MNRAS.288.1060B} reported that the reliable
SPH simulation of the could core collapse needs to resolve the local
Jeans mass, and requires $N_p \gg 10^{4}$.  Thus our current
resolution is significantly better than the criterion.

Molecular cloud cores are parameterized with two parameters $\alpha$
and $\gturb$.  Following \citet{1984ApJ...279..621M}, $\alpha$ is
defined as
%%%%%%%%%%%%%%%%%%%%%%%%%%%%%%
\begin{align}
  \label{eq:alpha}
  \alpha =\frac{E_{\rm thermal}}{|E_{\rm grav}|},
\end{align}
%%%%%%%%%%%%%%%%%%%%%%%%%%%%%%
where $E_{\rm thermal}=3c_{\rm s,0}^2 M_\odot/2$ and $E_{\rm grav}=
-3GM_{\odot}^2/5R_{\rm init}$ are the thermal and gravitational
energies corresponding to a homogeneous sphere of $1M_\odot$.  Then
the initial radius of the cloud core, $R_{\rm init}$ is written as
%%%%%%%%%%%%%%%%%%%%%%%%%%%%%%
\begin{align}
  \label{eq:Rinit}
  R_{\rm init} = \frac{2GM_{\odot}}{5c_{\rm s,0}^{2}} \alpha.
\end{align}
%%%%%%%%%%%%%%%%%%%%%%%%%%%%%%
The relative strength of the turbulence is parameterized by the virial
parameter $\gturb$ defined \citet{1992ApJ...395..140B}.  More
specifically, it is given by the ratio of the turbulence and
gravitational energies of the initial cloud:
%%%%%%%%%%%%%%%%%%%%%%%%%%%%%%
\begin{align}
  \label{eq:gamma_turb}
  \gturb = \frac{E_{\rm turb}}{|E_{\rm grav}|}
  = \frac{5\sigma^2_{\rm v} R_{\rm init}}{2GM_{\odot}},
\end{align}
%%%%%%%%%%%%%%%%%%%%%%%%%%%%%%
where $E_{\rm turb}=3\sigma^2_{\rm v}M_{\odot}/2$ with $\sigma_{\rm
  v}$ being the one-dimensional velocity dispersion of the turbulent
molecular cloud core.

%%%%%%%%%%%%%%%%%%%%%%%%%%%%%%
\begin{figure*}%%% Figure 1
  \includegraphics[clip,width=130mm]{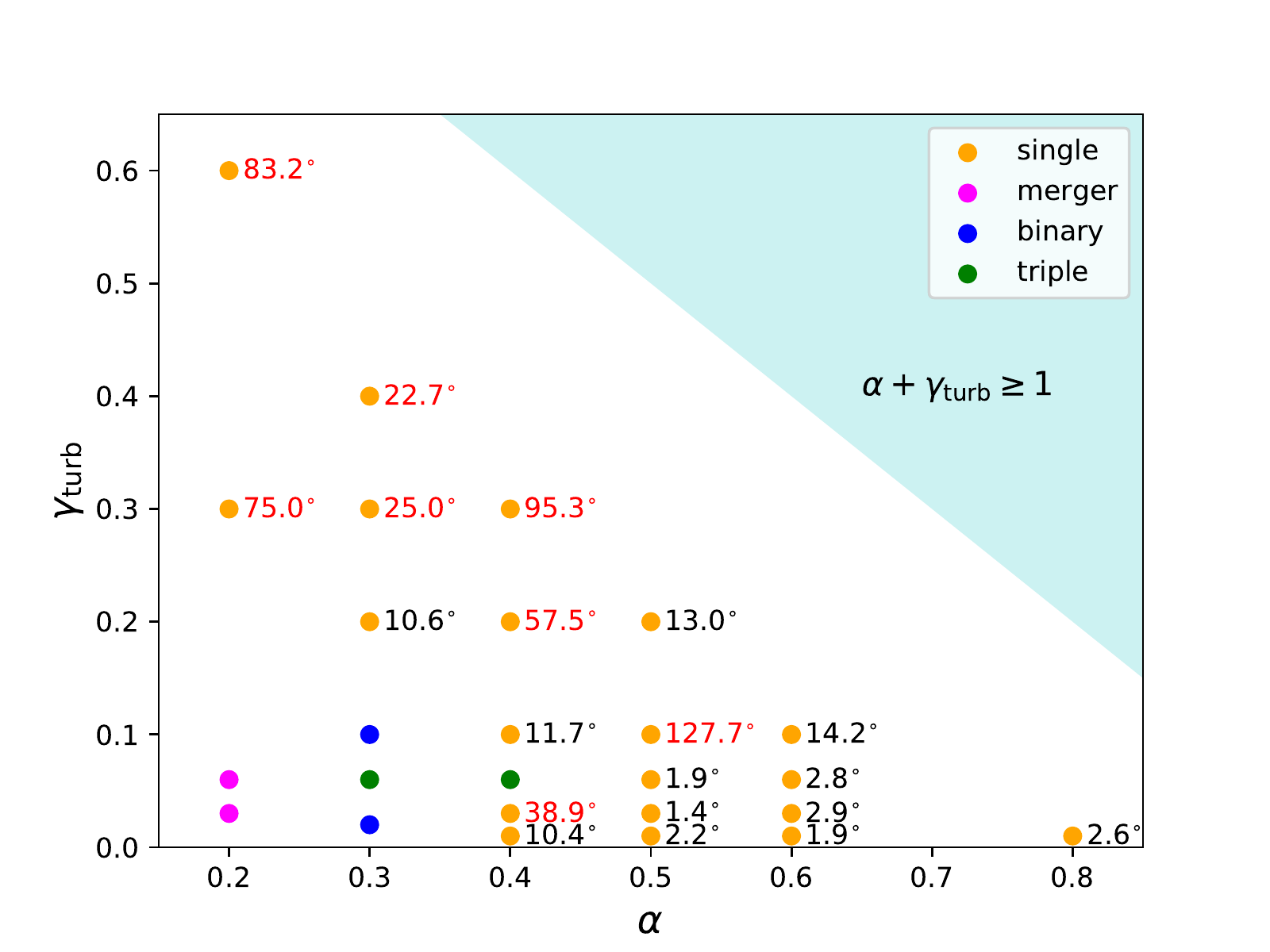}
  \caption{ ($\alpha$, $\gturb$) of the models.  The upper-right
    shaded region corresponds to gravitationally unbound cloud
    cores. The multiplicity of the protostars is shown with different
    colors.  The numbers at each point indicate the initial star-disk
    angle $\psd$ at the formation epoch of the sink particle for
    single star cases.  Models with $\psd > 20^{\circ}$ are indicated
    with red numbers.
    \label{fig:alpha-gamma}
  }
\end{figure*}
%%%%%%%%%%%%%%%%%%%%%%%%%%%%%%

%%%%%%%%%%%%%%%%%%%%%%%%%%%%%%
\begin{table*}%%% Table 1
  \centering
  \caption{
    The model names and parameters which characterize
    the initial molecular cloud core;
    $\alpha=E_{\rm thermal}/|E_{\rm grav}|$,
    $\gturb=E_{\rm turb}/|E_{\rm grav}|$,
    $\beff$ is the dimensionless angular momentum,
    $R_{\rm init}$ and $\rho_{\rm init}=3M_{\rm init}/(4\pi R_{\rm init}^3)$
    are the initial radius and density of the cloud cores,
    ${\cal M}$ is the mean Mach number,
    $t_{\rm ff}=\sqrt{3\pi/(32G\rho_{\rm init})}$ is the free-fall time of 
    the initial cloud cores,
    $\psd(t=10^2\rm{yr})$ and $\psd(t=10^5\rm{yr})$ are the
    star-disk angles measured at $t=10^2$ yr and $t=10^5$ yr from
    the protostar formation epoch.
    The last column indicates the multiplicity of the protostars in
    the simulation.  See subsection \ref{subsec:IC} for further
    detail.  
  }
  \begin{tabular}{ccccccccccc}
    \hline
    \hline
    Model & $\alpha$ & $\gturb$ & $\beff$ & 
    $R_{\rm init}~[\rm au]$ & $\rho_{\rm init}~[{\rm g~cm^{-3}}]$ & $\mach$ & $t_{\rm ff}~[\rm yr]$ &
    $\psd(t=10^2\rm{yr})$ & $\psd(t=10^5\rm{yr})$ & multiplicity \\
    \hline
    A1 & 0.2 & 0.03 & 0.0036 & 1967 & $1.9\times10^{-17}$ & 0.67 & $1.5\times10^{4}$ & $-$
    & $-$ & single(merger)\\
    A2 & 0.2 & 0.06 & 0.0072 & 1967 & $1.9\times10^{-17}$ & 0.95 & $1.5\times10^{4}$ & $-$
    & $-$ & single(merger)\\
    A3 & 0.2 & 0.3  & 0.036  & 1967 & $1.9\times10^{-17}$ & 2.1  & $1.5\times10^{4}$ &
    $75.0^{\circ}$ & $11.0^{\circ}$ & single \\
    A4 & 0.2 & 0.6  & 0.072  & 1967 & $1.9\times10^{-17}$ & 3.0  & $1.5\times10^{4}$ &
    $83.2^{\circ}$ & $8.6^{\circ}$ & single \\
    
    B1 & 0.3 & 0.02 & 0.0024 & 2950 & $5.5\times10^{-18}$ & 0.45 & $2.8\times10^{4}$ & $-$
    & $-$ & binary \\
    B2 & 0.3 & 0.06 & 0.0072 & 2950 & $5.5\times10^{-18}$ & 0.77 & $2.8\times10^{4}$ & $-$
    & $-$ & triple \\
    B3 & 0.3 & 0.1  & 0.012  & 2950 & $5.5\times10^{-18}$ & 1.0  & $2.8\times10^{4}$ & $-$
    & $-$ & binary \\
    B4 & 0.3 & 0.2  & 0.024  & 2950 & $5.5\times10^{-18}$ & 1.4  & $2.8\times10^{4}$ &
    $10.6^{\circ}$ & $10.0^{\circ}$ & single \\
    B5 & 0.3 & 0.3  & 0.036  & 2950 & $5.5\times10^{-18}$ & 1.7  & $2.8\times10^{4}$ &
    $25.0^{\circ}$ & $14.2^{\circ}$ & single \\
    B6 & 0.3 & 0.4  & 0.048  & 2950 & $5.5\times10^{-18}$ & 2.0  & $2.8\times10^{4}$ &
    $22.7^{\circ}$ & $10.2^{\circ}$ & single \\
    
    C1 & 0.4 & 0.01 & 0.0012 & 3933 & $2.3\times10^{-18}$ & 0.27 & $4.3\times10^{4}$ &
    $10.4^{\circ}$ & $5.2^{\circ}$ & single \\
    C2 & 0.4 & 0.03 & 0.0036 & 3933 & $2.3\times10^{-18}$ & 0.47 & $4.3\times10^{4}$ &
    $38.9^{\circ}$ & $3.9^{\circ}$ & single \\
    C3 & 0.4 & 0.06 & 0.0072 & 3933 & $2.3\times10^{-18}$ & 0.67 & $4.3\times10^{4}$ &  $-$
    &  $-$ & triple \\
    C4 & 0.4 & 0.1  & 0.012  & 3933 & $2.3\times10^{-18}$ & 0.87 & $4.3\times10^{4}$ & 
    $11.7^{\circ}$ & $6.0^{\circ}$ & single \\
    C5 & 0.4 & 0.2  & 0.024  & 3933 & $2.3\times10^{-18}$ & 1.2  & $4.3\times10^{4}$ &
    $57.5^{\circ}$ & $8.4^{\circ}$ & single \\
    C6 & 0.4 & 0.3  & 0.036  & 3933 & $2.3\times10^{-18}$ & 1.5  & $4.3\times10^{4}$ &
    $95.3^{\circ}$ & $7.2^{\circ}$ & single \\

    D1 & 0.5 & 0.01 & 0.0012 & 4917 & $1.2\times10^{-18}$ & 0.24 & $6.1\times10^{4}$ &
    $2.2^{\circ}$ & $2.5^{\circ}$ & single \\
    D2 & 0.5 & 0.03 & 0.0036 & 4917 & $1.2\times10^{-18}$ & 0.42 & $6.1\times10^{4}$ &
    $1.4^{\circ}$ & $5.6^{\circ}$ & single \\
    D3 & 0.5 & 0.06 & 0.0072 & 4917 & $1.2\times10^{-18}$ & 0.60 & $6.1\times10^{4}$ &
    $1.9^{\circ}$ & $6.5^{\circ}$ & single \\
    D4 & 0.5 & 0.1  & 0.012  & 4917 & $1.2\times10^{-18}$ & 0.77 & $6.1\times10^{4}$ &
    $127.7^{\circ}$ & $8.7^{\circ}$ & single \\
    D5 & 0.5 & 0.2  & 0.024  & 4917 & $1.2\times10^{-18}$ & 1.1  & $6.1\times10^{4}$ &
    $13.0^{\circ}$ & $14.2^{\circ}$ & single \\
    
    E1 & 0.6 & 0.01 & 0.0012 & 5900 & $6.9\times10^{-19}$ & 0.22 & $8.0\times10^{4}$ &
    $1.9^{\circ}$ & $6.8^{\circ}$ & single \\
    E2 & 0.6 & 0.03 & 0.0036 & 5900 & $6.9\times10^{-19}$ & 0.39 & $8.0\times10^{4}$ &
    $2.9^{\circ}$ & $5.2^{\circ}$ & single \\
    E3 & 0.6 & 0.06 & 0.0072 & 5900 & $6.9\times10^{-19}$ & 0.55 & $8.0\times10^{4}$ &
    $2.8^{\circ}$ & $7.8^{\circ}$ & single \\
    E4 & 0.6 & 0.1  & 0.012  & 5900 & $6.9\times10^{-19}$ & 0.71 & $8.0\times10^{4}$ &
    $14.2^{\circ}$ & $7.5^{\circ}$ & single \\

    F1 & 0.8 & 0.01 & 0.0012 & 7866 & $2.9\times10^{-19}$ & 0.19 & $1.2\times10^{5}$ &
    $2.6^{\circ}$ & $1.2^{\circ}$ & single \\
    \hline
    \hline
  \end{tabular}
  \label{tb:initial_conditions}
  \footnotesize
\end{table*}
%%%%%%%%%%%%%%%%%%%%%%%%%%%%%%

We consider 26 models specified by the different set of $\alpha$ and
$\gturb$ (see Figure \ref{fig:alpha-gamma} and Table
\ref{tb:initial_conditions}).  We impose $\alpha+\gturb \lesssim 0.8$
because the cloud cores are supposed to be nearly virialized in
reality. We do not assign the angular momentum of the initial core
{\it a priori}.  Due to the stochastic nature of the turbulent
velocity field, however, the core acquires a non-vanishing net angular
momentum $\bm{J}_{\rm init}$.  Thus we set the direction of
$\bm{J}_{\rm init}$ as the $z$-axis of each simulation model.

Table \ref{tb:initial_conditions} lists the dimensionless angular
momentum of the core:
%%%%%%%%%%%%%%%%%%%%%%%%%%%%%%
\begin{align}
  \label{eq:beff}
  \beff = \frac{25}{12}
  \frac{ |\bm{J}_{\rm init}|^{2} }{ G M_\odot^{3} R_{\rm init} },
\end{align}
%%%%%%%%%%%%%%%%%%%%%%%%%%%%%%
and other parameters for each model:
%%%%%%%%%%%%%%%%%%%%%%%%%%%%%%
\begin{align}
  \label{eq:rho_init}
  \rho_{\rm init} &= \frac{3M_{\odot}}{4\pi R_{\rm init}^3} , \\
  \label{eq:t_ff}
  t_{\rm ff} &= \sqrt{\frac{3\pi}{32G\rho_{\rm init}}} , \\
  \label{eq:mach_number}
  {\cal M} &= \frac{1}{c_{\rm s,0}}
  \sqrt{\frac{1}{N_{\rm p}}\sum_{i=1}^{N_{\rm p}} v^2_{i,\rm{SPH}}} ,
\end{align}
%%%%%%%%%%%%%%%%%%%%%%%%%%%%%%
where $\rho_{\rm init}$, $t_{\rm ff}$, and ${\cal M}$ are the density,
free-fall time, and mean Mach number of the initial cloud core,
respectively.  The last column of Table \ref{tb:initial_conditions}
indicates the multiplicity of the protostars formed at the end of our
simulation $\sim 10^{5}$ yr.
%%%%%%%%%%%%%%%%%%%%%%%%%%%%%%%%%%%%%%%%%%%%%%%%%%%%%%%%%%%%%

%%%%%%%%%%%%%%%%%%%%%%%%%%%%%%
\begin{figure*}%%% Figure 2
  \includegraphics[clip,width=140mm]{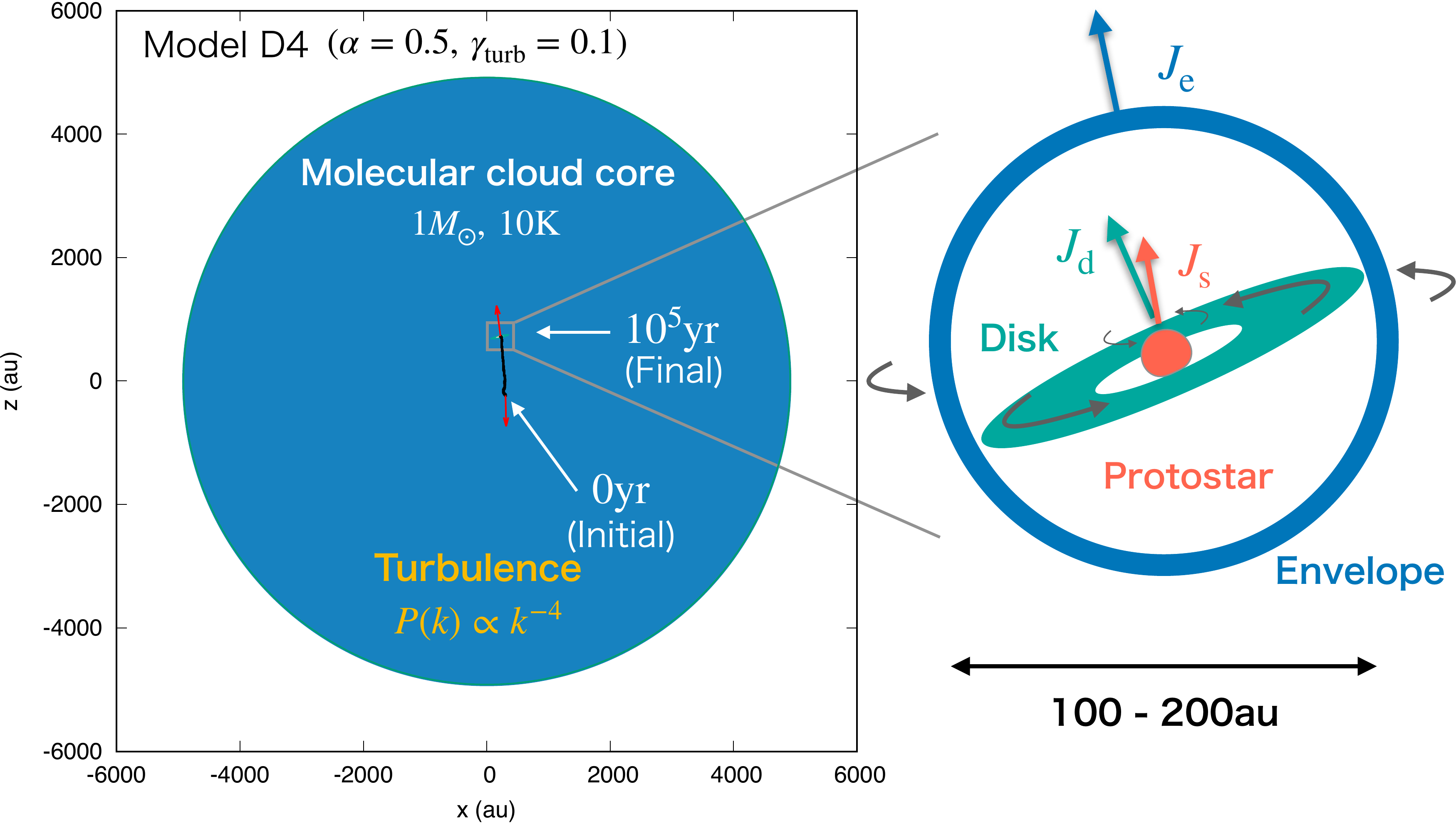}
  \caption{ Schematic illustration of the configuration of the initial
    cloud core, circumstellar envelope, protoplanetary disk and
    protostar in model D4.  $\Js$, $\Jd$ and $\Je$ are the angular
    momentum of the protostar spin, protoplanetary disk rotation and
    circumstellar envelope rotation, respectively.  Two red arrows in
    the left figure show $\Js$ at $t=0$ yr and $t=10^5$ yr.
    \label{fig:Schematic-D4}
  }
\end{figure*}
%%%%%%%%%%%%%%%%%%%%%%%%%%%%%%

%%% 2.3. definiton of angles
\subsection{
  Definitions of Protostar, Disk and Envelope in our Simulation
}
\label{subsec:star-disk-env}
%%%%%%%%%%%%%%%%%%%%%%%%%%%%%%%%%%%%%%%%%%%%%%%%%%%%%%%%%%%%%

The estimate of the star-disk angle crucially depends on the
definition of the protostar and disk in the simulations. Our SPH
simulation resolves the spatial structure of the disk very well, but
not the protostar {\it at all}.  Instead, we adopt a sink particle
technique to identify a 1 au sphere enclosing the protostar.  The
mass, velocity and angular momentum of the sink particle, $M_{\rm
  s}(t)$, $\bm{v}_{\rm s}(t)$ and $\bm{J}_{\rm s}(t)$, can be directly
computed from simulations.  Nevertheless they are not identical to
those of the protostar that is supposed to occupy merely the central
$\sim 5 \times 10^{-3}$ au scale.  In particular, it is well known
that a substantial fraction of $\bm{J}_{\rm s}(t)$ should be removed
from the region since it would exceed the breakup value of the stellar
surface otherwise.

Indeed, several numerical schemes have been proposed to empirically
limit the amount of the angular momentum accreted onto the central
protostar \citep{2010MNRAS.401.1505B, 2015MNRAS.450.3306F}.  For
instance, \citet{2010MNRAS.401.1505B} were interested in the
reorientation channel of the inner disk and the central protostar by
the warp propagation, and assumed that the protostar (sink particle)
acquires the mass and angular momentum transported through the
protoplanetary disk alone.  \citet{2015MNRAS.450.3306F}, on the other
hand, did not allow that the accreted angular momentum exceeds the
breakup value of the stellar spin, since they were interested in the
star-disk alignment mechanism due to the gravitational torque between
the spin-induced stellar quadrupole and the surrounding disk.

As briefly mentioned in the previous section, we re-simulated one of
the model following the sub-grid procedure of
\citet{2015MNRAS.450.3306F}.  We made sure that the final star-disk
angle $\psd$ is well converged to the value without implementing the
procedure, while the amplitude of the stellar spin is significantly
suppressed.  In addition, as shown by \citet{2018MNRAS.475.5618B}, the
star-disk misalignment can be captured even without the sub-grid model
in the calculation of the star cluster formation.  Therefore, we do
not introduce the sub-grid model in what follows.

We define the protoplanetary disk as a set of SPH particles that
satisfy the following criteria:
%%%%%%%%%%%%%%%%%%%%%%%%%%%%%%%%%%%%%%%%%%%%%%%%%%%%%%
\begin{align}
  \label{eq:disk-bound}
  \frac{1}{2}{v}_{i, \rm{SPH}}^2 -
  \frac{GM_{\rm s}}{|\bm{r}_{i, {\rm SPH}} - \bm{r}_{\rm s}|} &< 0, \\
  \label{eq:disk-virial}
  2 ~ |(\bm{v}_{i, \rm{SPH}}-\bm{v}_{\rm s})_{r}| &<
  |(\bm{v}_{i, \rm{SPH}}-\bm{v}_{\rm s})_{t}|, \\
  \label{eq:disk-maxsize}
  |\bm{r}_{i, {\rm SPH}} - \bm{r}_{\rm s}| &< r_{\rm limit} = \rm{500~au},
\end{align}
%%%%%%%%%%%%%%%%%%%%%%%%%%%%%%%%%%%%%%%%%%%%%%%%%%%%%%
where the subscripts $r$ and $t$ denoting the radial and tangential
components of the relative velocity.  Equation (\ref{eq:disk-bound})
checks whether the SPH particles is bound to a sink particle.
Equation (\ref{eq:disk-virial}), which checks whether the rotation of
the SPH particle is much faster than the infall, is introduced to
define the rotation plane of the disk more precisely.  The disk
rotation axis fluctuates without this condition.  Equation
(\ref{eq:disk-maxsize}) introduces the maximum size of the disk.  We
confirm, however, that the real size of the disk is determined by
equations (\ref{eq:disk-bound}) and (\ref{eq:disk-virial}), and our
result is not changed by the choice of $r_{\rm limit}$ between 200 to
500 au.

Finally we define the circumstellar envelope surrounding the
protoplanetary disk.  In this paper, the envelope is defined as a set
of all the SPH particles within 2000 au from the sink particle.
Hence, the envelope also includes the disk gas.

Adopting the above definitions of the protostar, protoplanetary disk,
and circumstellar envelope, we compute the angular momenta of the
protostar spin, protoplanetary disk rotation, and circumstellar
envelope rotation, $\Js$, $\Jd$, and $\Je$, at each epoch.  The
relative angles between $\Js$, $\Jd$, and $\Je$ are defined as
%%%%%%%%%%%%%%%%%%%%%%%%%%%%%%
\begin{align}
  \label{eq:psd}
  \psd &= \cos^{-1} \left( \frac{\Js \cdot \Jd }{|\Js| |\Jd|} \right), \\
  \label{eq:pse}
  \pse &= \cos^{-1} \left( \frac{\Js \cdot \Je }{|\Js| |\Je|} \right), \\
  \label{eq:pde}
  \pde &= \cos^{-1} \left( \frac{\Jd \cdot \Je }{|\Jd| |\Je|} \right).
\end{align}
%%%%%%%%%%%%%%%%%%%%%%%%%%%%%%
We mainly investigate the time evolution of these angles in this
paper.

%%%%%%%%%%%%%%%%%%%%%%%%%%%%%%%%%%%%%%%%%%%%%%%%%%%%%%%%%%%%%

%%% 3. results of fiducial model
\section{Evolution of $\psd$, $\pse$ and $\pde$ for our Fiducial Model}
\label{sec:D4-result}
%%%%%%%%%%%%%%%%%%%%%%%%%%%%%%%%%%%%%%%%%%%%%%%%%%%%%%%%%%%%%%%%%%

Before proceeding to the statistical analysis, we focus on model D4
that represents a virialized ($\alpha=0.5$) and reasonably strong but
still subsonic turbulence (${\cal M}=0.77$).  Thus we adopt this model
as our fiducial example, and discuss its detailed evolutionary
behavior in this section.

Figure \ref{fig:Schematic-D4} schematically illustrates the
configuration of our simulation result for model \FiducialModel.  The
simulation starts from the molecular cloud core with the radius of
$\sim 5000$ au.  The molecular cloud core immediately gravitationally
collapses, and the sink particle forms close to the center of the
initial cloud core.  We define the origin of the time $t=0$ as at the
formation epoch of the protostar.  The spin of the sink particle shown
with red arrow is initially almost anti-parallel to the $z$-axis.  The
trajectory of the sink particle during the evolution is indicated by
the black curve, and its spin shown with red arrow becomes aligned
after $10^5$ yr.  The close-up schematic figure of the system in
Figure \ref{fig:Schematic-D4} shows the configuration of the angular
momentum of the protostar, protoplanetary disk, and circumstellar
envelope at $t=10^5$ yr.

%%%%%%%%%%%%%%%%%%%%%%%%%%%%%%
\begin{figure*}%%% Figure 3 and Figure 4
  \centering
  \includegraphics[clip,width=100mm,angle=-90]{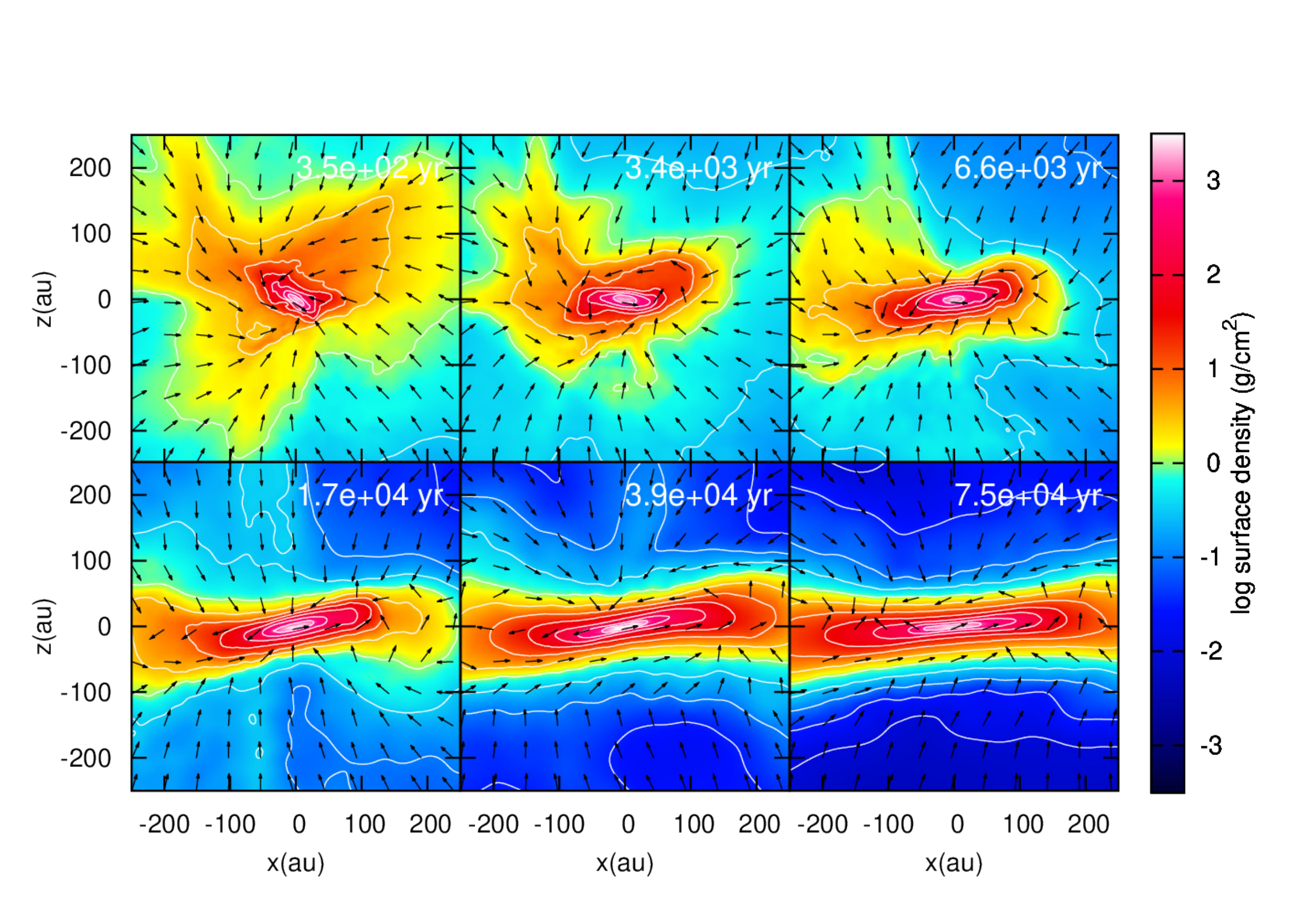}
  \caption{ The surface density evolution on the x-z plane for model
    \FiducialModel.  The top-left, top-middle and top-right panels
    show snapshots at $3.5\times10^2$ yr, $3.4\times10^3$ yr and
    $6.6\times10^3$ yr, respectively.  The bottom-left, bottom-middle
    and bottom-right panels show snapshots at $1.7\times10^4$ yr,
    $3.9\times10^4$ yr and $7.5\times10^4$ yr, respectively.  White
    lines show contours of the surface density.  Black arrows show
    directions of the density weighted velocity.
    \label{fig:snapshots_xz_of_fiducial_model}
  }
  \includegraphics[clip,width=100mm,angle=-90]{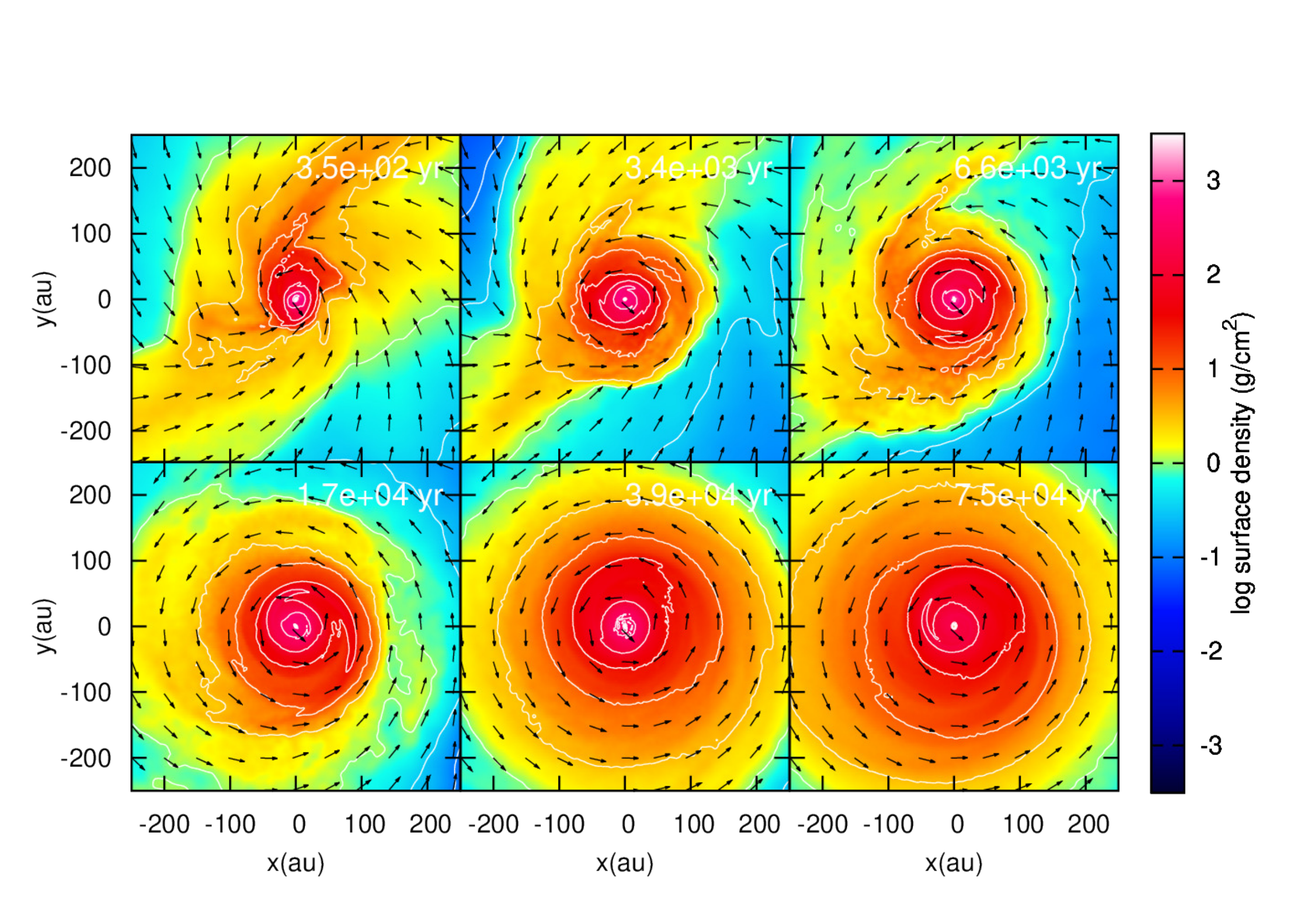}
  \caption{ Same as Figure \ref{fig:snapshots_xz_of_fiducial_model}
    but on $x$-$y$ plane.
    \label{fig:snapshots_xy_of_fiducial_model}
  }
\end{figure*}
%%%%%%%%%%%%%%%%%%%%%%%%%%%%%%

Figures \ref{fig:snapshots_xz_of_fiducial_model} and
\ref{fig:snapshots_xy_of_fiducial_model} shows the surface density
evolution on $x$-$z$ and $x$-$y$ planes.  The position of the sink
particle is fixed at the center.  We define the formation epoch of the
sink particle as the origin of the time.

Figure \ref{fig:snapshots_xz_of_fiducial_model} indicates that, just
after the protostar formation, the surface density has the dense
filamentary structure (top left panel).  As time proceeds, the
coherent disk structure develops.  Coincidentally, the filamentary
structure disappears.  This indicates that the mass accretion to the
protostar is random at the protostar formation epoch and is mainly
from the protoplanetary disk in the later phase.

We also find that the disk rotation axis is gradually changing during
the evolution.  At the early phase, the disk rotation axis is tilted
from $z$ axis (e.g., top middle panel of Figure
\ref{fig:snapshots_xz_of_fiducial_model}).  In subsequent evolution,
it gradually becomes aligned to the $z$-axis.  The spiral arms are
formed in the bottom-left and bottom-middle panels of Figure
\ref{fig:snapshots_xy_of_fiducial_model}, which is caused by the
gravitational instability.

%%%%%%%%%%%%%%%%%%%%%%%%%%%%%%
\begin{figure*}%%% Figure 5
  \centering
  \includegraphics[clip,width=100mm,angle=-90]{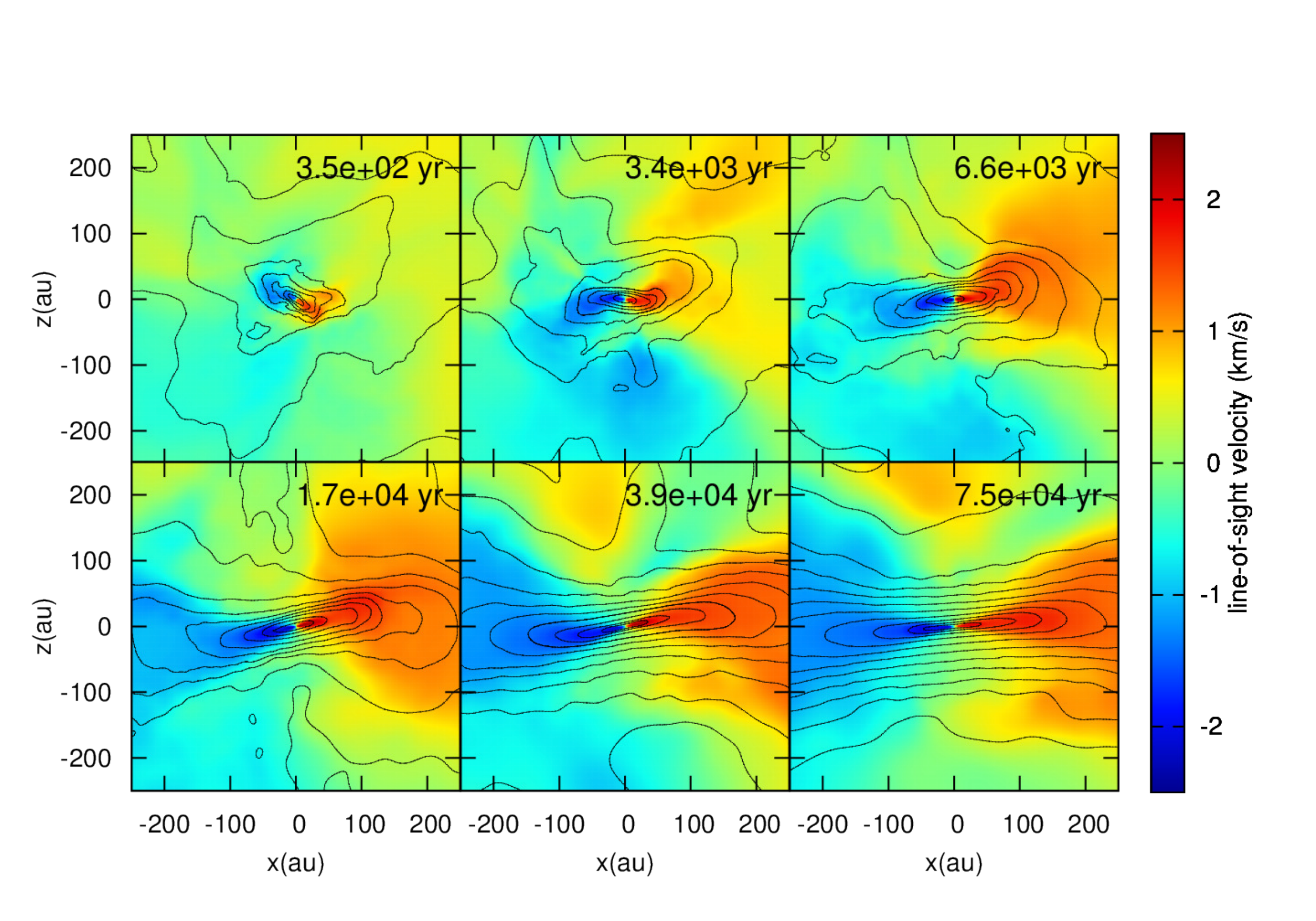}
  \caption{ The evolution of the density weighted line-of-sight
    velocity on the x-z plane for model \FiducialModel.  The epochs of
    each panel are the same as Figure
    \ref{fig:snapshots_xz_of_fiducial_model}.  Black lines are
    contours of the surface density.
    \label{fig:Moment1_0501}
  }
\end{figure*}
%%%%%%%%%%%%%%%%%%%%%%%%%%%%%%

Figure \ref{fig:Moment1_0501} shows the density weighted line-of-sight
velocity along the $y$-axis.  At the formation epoch of the protostar
(top-left panel), the regions with positive (red) and negative
velocity (blue) is mixed around the protostar, indicating that the
turbulent velocity field is maintained around the protostar.  As time
proceeds, the structure of the rotation becomes coherent, indicating
that the rotationally supported disk develops.  We note that the
filamentary structure in the top panels of Figure
\ref{fig:snapshots_xz_of_fiducial_model} is infalling and not
outflowing.

%%%%%%%%%%%%%%%%%%%%%%%%%%%%%%
\begin{figure}%%% Figure 6
  \centering
  \includegraphics[clip,width=80mm,angle=0]{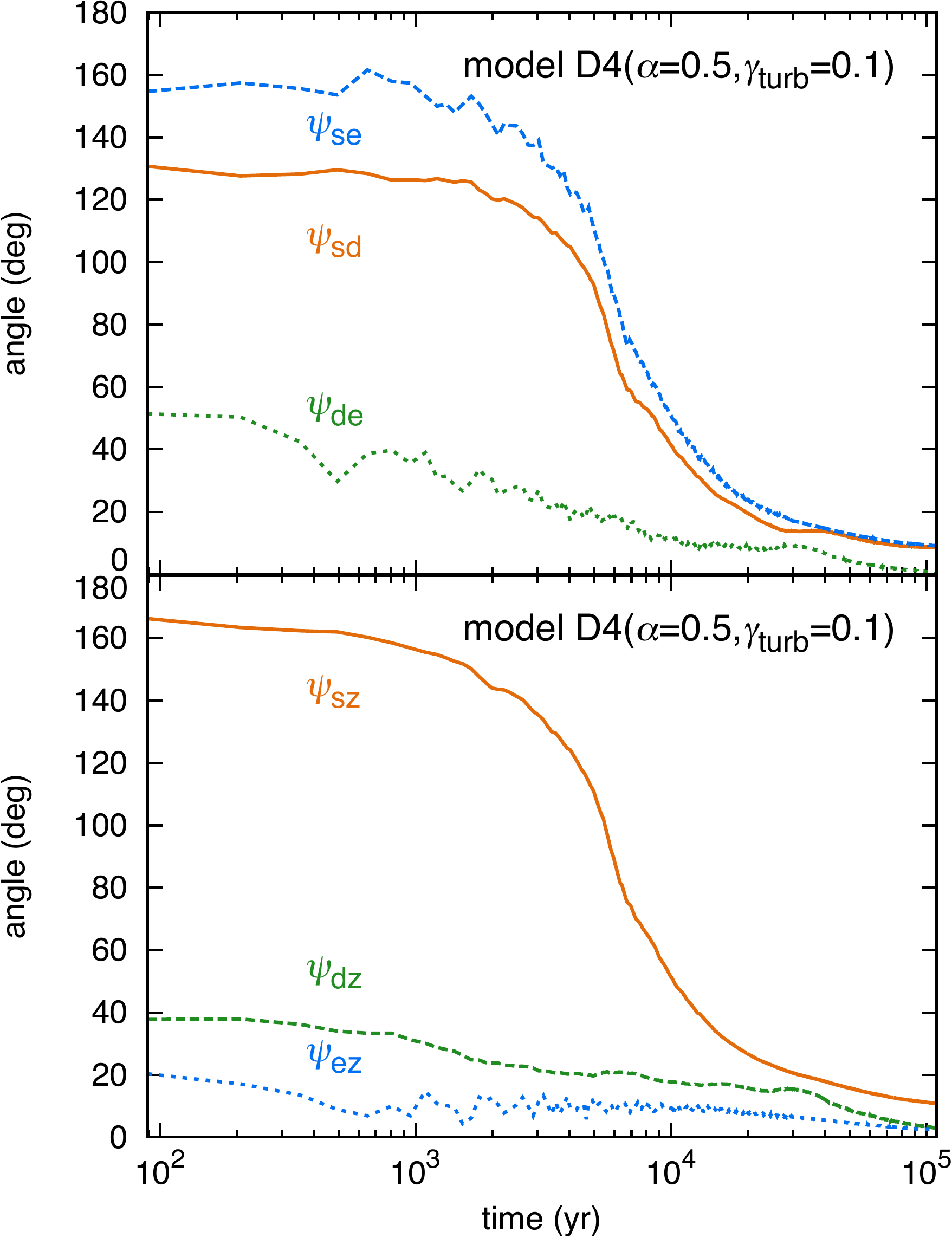}
  \caption{ Time evolution of the directions of the stellar spin, disk
    rotation and envelope rotation for model \FiducialModel.  The
    red-solid, blue-dashed and green-dotted lines in the upper panel
    show star-disk relative angle $\psd(t)$, star-envelope relative
    angle $\pse(t)$, and disk-envelope relative angle $\pde(t)$,
    respectively.  The red-solid, green-dashed and blue-dotted lines
    show the stellar spin angle $\psi_{\rm sz}(t)$, disk rotation
    angle $\psi_{\rm dz}(t)$ envelope rotation angle $\psi_{\rm
      ez}(t)$ from $z$ axis.
    \label{fig:angles_D4}
  }
\end{figure}
%%%%%%%%%%%%%%%%%%%%%%%%%%%%%%

Figure \ref{fig:angles_D4} shows the time evolution of the directions
of the stellar spin, disk rotation and envelope rotation.  Because the
disk surrounding the protostar becomes well developed $\sim10^2$ yr
after the protostar formation, we plot $\psd(t)$, $\pse(t)$, and
$\pde(t)$ for $10^2~{\rm yr}<t<10^5~{\rm yr}$.  The lower panel shows
the angles of the stellar spin, disk rotation, and envelope rotation
axes relative to the $z$-axis; $\psi_{\rm sz}(t)$, $\psi_{\rm dz}(t)$,
and $\psi_{\rm ez}(t)$.

By the filamentary mass accretion toward the center in the early
evolution phase (top left panel of Figure
\ref{fig:snapshots_xz_of_fiducial_model}), the $\psi_{\rm sz}$ at the
early formation epoch has the large value of $\psi_{\rm sz}\sim
150^\circ$ and the stellar spin is significantly different from the
rotation direction of the initial cloud core ($z$-axis).  On the other
hand, the larger-scale gas distribution shares the initial cloud core
rotation and $\psi_{\rm dz}$ and $\psi_{\rm ez}$\ are already small
even at $t<10^3$ yr.  As a results, the relative angle between the
protostar spin and the protoplanetary disk rotation $\psd$ or the
envelope rotation $\pse$ also have large values of $\gtrsim 120^\circ$
meaning that the protostar spin and disk rotation or envelope rotation
are highly misaligned.

In the subsequent evolution phase, the protostar spin evolves mainly
by the accretion of the angular momentum from the disk, and the $\psd$
begins to decrease in $t>10^3$ yr.  Simultaneously, the disk angular
momentum evolves via the accretion of the angular momentum from the
envelope, and the $\pde$ decreases.

Because of the angular momentum conservation of the entire system,
$\psi_{\rm sz}$, $\psi_{\rm dz}$, $\psi_{\rm ez}$ becomes $\sim
0^{\circ}$ at $t=10^5$ yr.  All rotation axes eventually align toward
the $z$-axis.  Note that $\psd$ significantly decreases at $t\sim
10^4$ yr.  As we will see bellow, this timescale corresponds to the
timescale in which the protostar forget its initial spin angular
momentum.

%%%%%%%%%%%%%%%%%%%%%%%%%%%%%%
\begin{figure}%%% Figure 7
  \centering
  \includegraphics[clip,width=155mm,angle=-90]{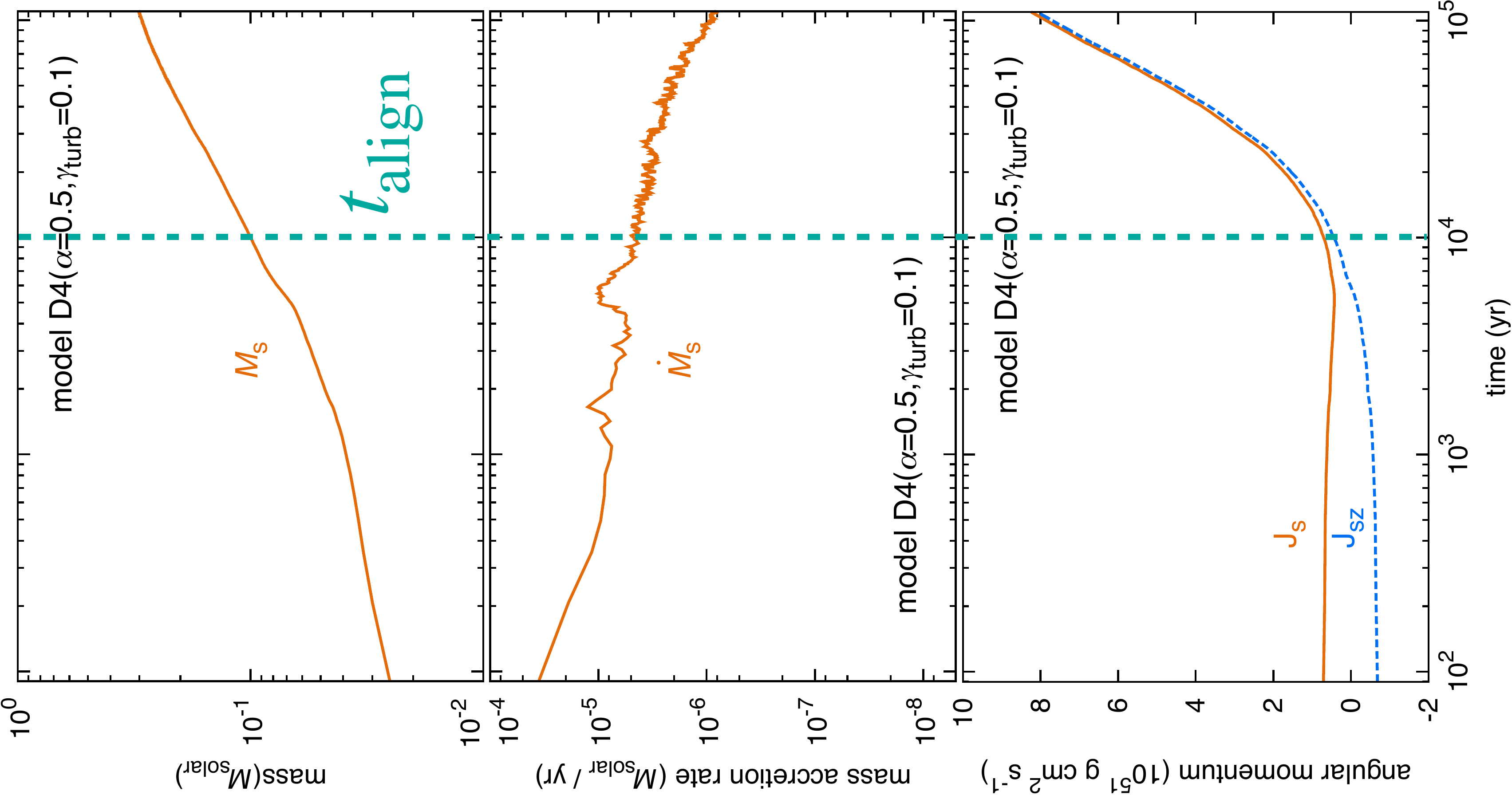}
  \caption{ Time evolution of the mass of the protostar (top), mass
    accretion rate onto the protostar (middle) and spin angular
    momentum of the protostar (bottom), respectively, for model
    \FiducialModel.  The horizontal axis shows the time from the
    protostar formation.
    \label{fig:time_mass_and_acc_rate_on_fiducial}
  }
\end{figure}
%%%%%%%%%%%%%%%%%%%%%%%%%%%%%%

Here, we show that the timescale of the alignment found in Figure
\ref{fig:snapshots_xz_of_fiducial_model} corresponds to the mass
increase timescale of the protostar.  Figure
\ref{fig:time_mass_and_acc_rate_on_fiducial} shows the time evolution
of the protostar mass $M_{\rm s}$ (top), the mass accretion rate of
the protostar ${\dot M}_{\rm s}$ (middle), and the angular momentum of
the protostar $J_{\rm s}=|\bm{J}_{\rm s}|$ (bottom).  At $t=10^2$ yr,
the protostar mass is $M_{\rm s}\sim 0.03\msun$, which is consistent
with the Jeans mass of the first core
\citep[e.g.,][]{2010ApJ...724.1006M}.  Subsequently, the protostar
mass increases by the mass accretion from the disk.  As shown in
middle panel, the mass accretion rate onto the protostar in $t<10^4$
yr is $\sim 10^{-5} M_\odot yr^{-1}$.  Thus, the protostar mass
increases by a factor of three in $\sim 10^4$ yr.  This means that the
timescale of $10^4$ yr corresponds to the mass growth timescale of the
protostar.

The bottom panel shows that $J_{\rm s}$ is almost constant in $t<10^4$
yr.  This indicates that the angular momentum supplied by the disk is
smaller than the inherent angular momentum of the protostar obtained
at its formation.  In $t>10^4$ yr, on the other hand, the angular
momentum supplied from the disk dominates the inherent angular
momentum of the protostar, meaning that the protostar forgets the
initial angular momentum.  Note that the stellar radius does not
change significantly during the protostar evolution phase, the angular
momentum accretion rate is proportional to the mass accretion rate.
Thus, we conclude that the alignment timescale of the stellar spin
corresponds to the mass growth timescale of the protostar.

With the consideration above, the characteristic timescale of the
alignment $t_{\rm align}$ can be estimated as
%%%%%%%%%%%%%%%%%%%%%%%%%%%%%%
\begin{align}
  \label{eq:t_align}
  t_{\rm align} = \frac{\varepsilon M_{\rm 0}}{\dot{M_{\rm s}}}
  \sim 10^{4}{\rm yr}
  \Biggl( \frac{M_{\rm 0}}{3\times 10^{-2} \msun} \Biggr)
  \Biggl(\frac{\dot{M_{\rm s}}}{10^{-5} \msun~{\rm yr}^{-1}} \Biggr)^{-1}
  \Biggl( \frac{\varepsilon}{3} \Biggr),
\end{align}
%%%%%%%%%%%%%%%%%%%%%%%%%%%%%%
where $\varepsilon \sim 2-3$ is an empirical fudge factor.  As we will
show in \S \ref{sec:statistical-result}, the timescale of equation
(\ref{eq:t_align}) well describes the alignment timescale of other
models.

Note that the mass accretion rate of $\dot{M}_{\rm s}$ in the early
evolution phase is highly uncertain and the smaller mass accretion
rate may realize.  If $\dot{M}_{\rm s}$ in the real molecular cloud
core is smaller than our simulations, it causes longer $t_{\rm
  align}$.  For example, if we take the smaller mass accretion rate of
$\dot{M}_{\rm s} \sim 10^{-6} \msun~\rm{yr}^{-1}$ as suggested by
\citet{1977ApJ...214..488S} and from the recent observations of Class
0/I Young Stellar Objects (YSOs) \citep[e.g.,][]{2017ApJ...834..178Y},
the alignment timescale $t_{\rm align}$ increases by a factor of 10,
and becomes $\sim 10^5$ yr.

%%%%%%%%%%%%%%%%%%%%%%%%%%%%%%%%%%%%%%%%%%%%%%%%%%%%%%%%%%%%%%%%%%

%%% 4. results of all models
\section{Statistical Analysis of $\psd$, $\pse$ and $\pde$
  and their Dependence on the Model Parameters}
\label{sec:statistical-result}
%%%%%%%%%%%%%%%%%%%%%%%%%%%%%%%%%%%%%%%%%%%%%%%%%%%%%%%%%%%%%%%%%%

The last column of Table \ref{tb:initial_conditions} indicates the
multiplicity of the protostars in each model at $\sim 10^{5}$ yr after
the formation of the first sink particle.

In models A1 and A2, two sink particles are formed, and they are
merged.  In models B1 and B3, a binary system with the separation of
$\sim30$ au and $\sim50$ au are formed, respectively.  Models B2 and
C3 correspond to the triple star formation cases, in which the binary
system formed at first and circumbinary disk rotates around them
changed into the third object due to the gravitational instability.

Those systems exhibit their own specific but interesting evolution
history, and we omit to discuss these results in this paper.  Thus, we
consider the remaining 20 models in this section.

%%%%%%%%%%%%%%%%%%%%%%%%%%%%%%
\begin{figure*}%%% Figure 8
  \centering
  \includegraphics[clip,width=130mm,angle=0]{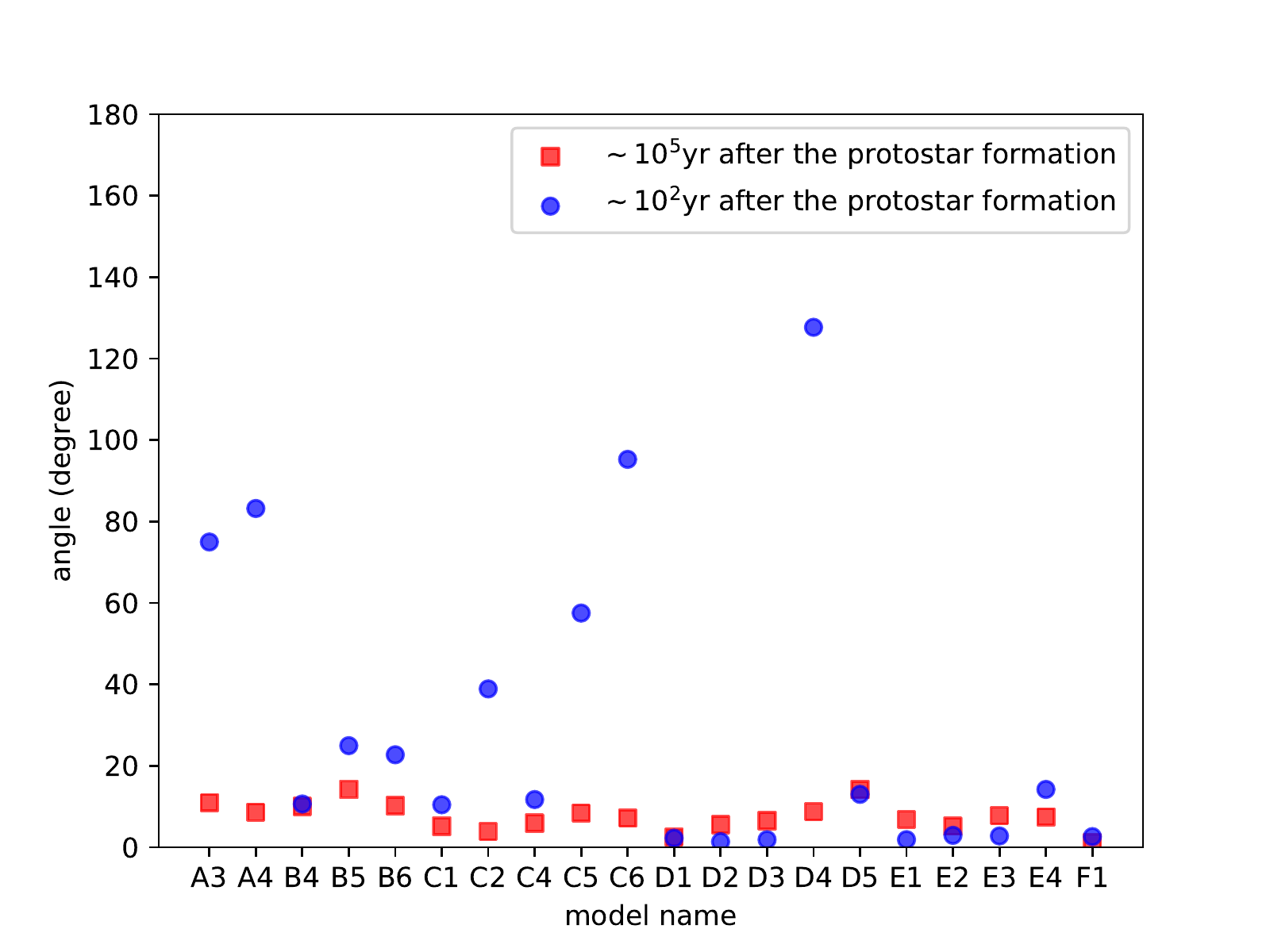}
  \caption{ $\psd$ at $t \sim 10^2$ yr (blue) and $t \sim 10^5$ yr
    (red) of different models.
    \label{fig:summary_psd}
  }
\end{figure*}
%%%%%%%%%%%%%%%%%%%%%%%%%%%%%%

Figure \ref{fig:summary_psd} summarizes the initial ($t \sim 10^2$
yr;blue) and final ($t \sim 10^5$ yr;red) values of $\psd$ for 20
models in which the protostar is formed as a single star.  Out of the
20 models, 12 models are aligned initially with $\psd<20^{\circ}$, and
the remaining 8 models are misaligned with $\psd>20^{\circ}$.  Figure
\ref{fig:alpha-gamma} indicates that $\psd$ is barely correlated with
$\alpha$ and $\gturb$.  This is because the initial $\psd$ is
determined by the local density and velocity fluctuation around the
sink particle, while $\alpha$ and $\gturb$ characterize the global
properties of the entire cloud core.  Nevertheless we may recognize a
weak positive trend of initial $\psd$ and $\gturb$ in Figures
\ref{fig:alpha-gamma} and \ref{fig:summary_psd}.  On the other hand,
the $\psd$ is $0^{\circ}<\psd<20^{\circ}$ at $10^5$ yr after the
protostar formation, independently of their initial values.

%%%%%%%%%%%%%%%%%%%%%%%%%%%%%%
\begin{figure}%%% Figure 9
  \centering
  \includegraphics[clip,width=160mm,angle=-90]{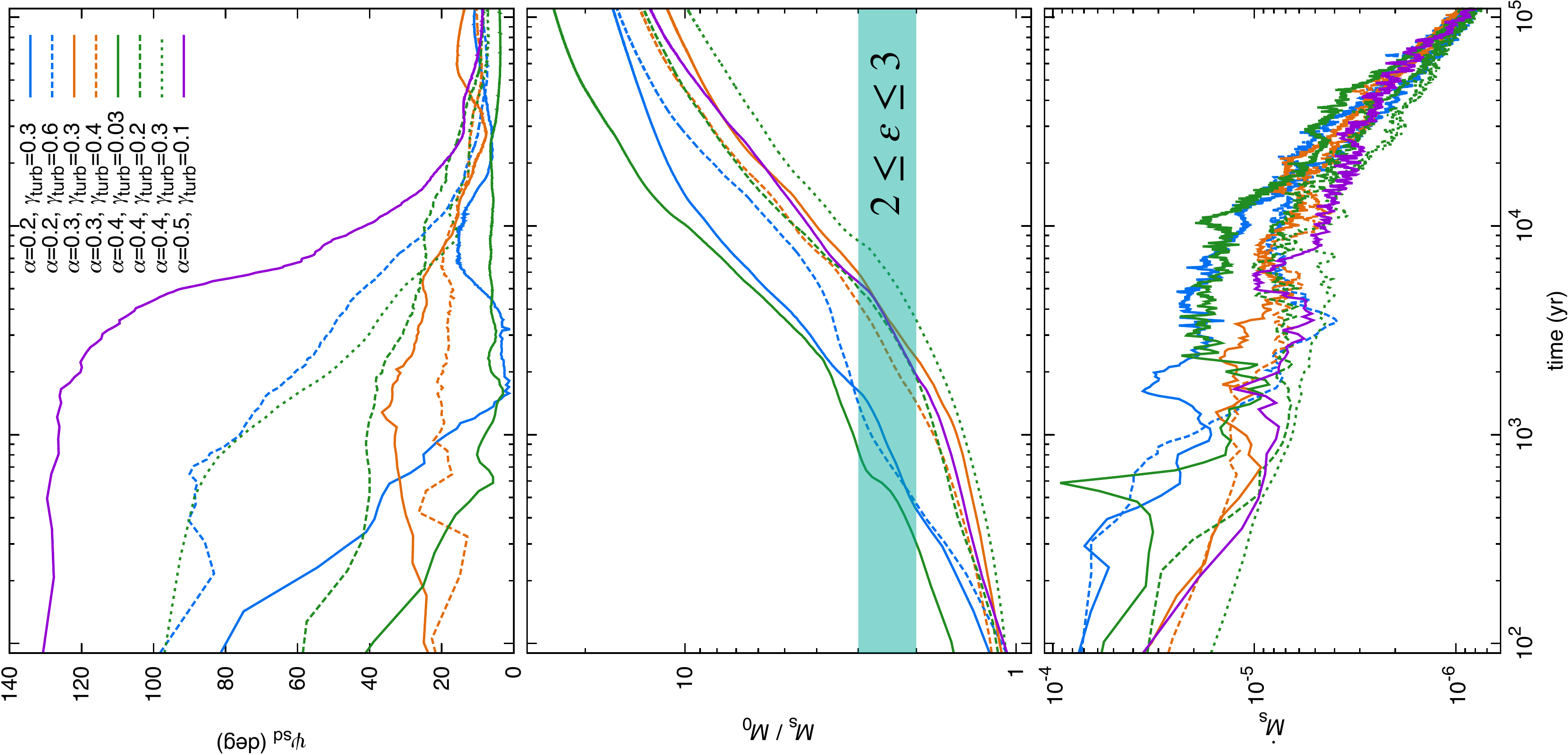}
  \caption{ Time evolution of $\psd$ (top), protostar mass (middle)
    and protostar mass accretion rate (bottom) of all simulations in
    which the star-disk misalignment appeared.  The horizontal axis of
    this plot correspond to the time from the protostar formation.
    This is the case that the protostar is formed as a single star.
    \label{fig:results_angles_all}
  }
\end{figure}
%%%%%%%%%%%%%%%%%%%%%%%%%%%%%%

Figure \ref{fig:results_angles_all} shows the evolution of $\psd$,
$M_{\rm s}$ and $\dot{M}_{\rm s}$ of the initially misaligned 8
systems ($\psd>20^{\circ}$ at $t\sim10^2$ yr).  The top panel shows
that the initial values of $\psd$ are distributed in
$20^{\circ}<\psd<130^{\circ}$, and they decrease to $\psd<20^{\circ}$
in the timescale of several $10^3$ yr to $10^4$ yr.

The middle panel of Figure \ref{fig:results_angles_all} shows that the
mass increase timescale varies from $\lesssim 10^3$ yr (green solid
line) to $\sim 10^4$ yr (green dotted line).  As expected from the
equation (\ref{eq:t_align}), $\psd$ of the model with the small mass
increase timescale (e.g., green solid line of the middle panel of
Figure \ref{fig:results_angles_all}) quickly decreases to
$\psd<10^\circ$ in $t<10^3$ yr.  The correlation between the small
mass increase timescale and the small alignment timescale suggests
that the equation (\ref{eq:t_align}) is a good estimate of the
alignment timescale of the stellar spin and disk rotation direction.
In all models considered in this paper, the final values of $\psd$
range from a few to $10^{\circ}$ and very small.

%%%%%%%%%%%%%%%%%%%%%%%%%%%%%%
\begin{figure}%%% Figure 10
  \centering
  \includegraphics[clip,width=60mm,angle=-90]{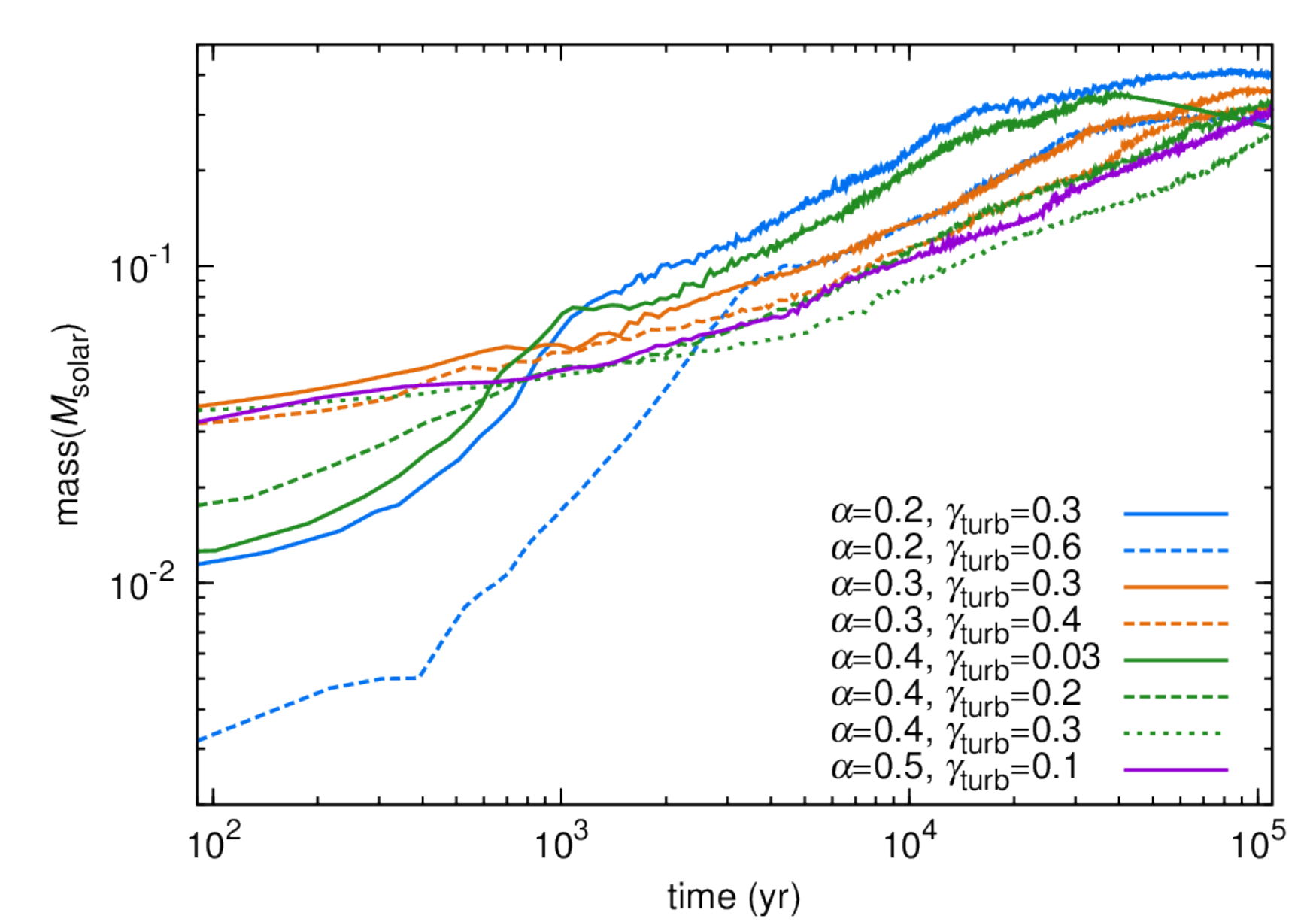}
  \caption{ Time evolution of the mass of the protoplanetary disk.  We
    plot the results of our simulations with star-disk misalignment.
    The horizontal axis of this plot shows the time after the
    protostar formation.
    \label{fig:disk_mass_all}
  }
\end{figure}
%%%%%%%%%%%%%%%%%%%%%%%%%%%%%%

Our current simulations predict relatively well-aligned star-disk
systems.  We note, however, a few cations here before drawing general
conclusions.  Firstly, we focus on the 20 single star systems, and do
not discuss the other six multiple-star systems out of the 26 models
summarized in Table \ref{tb:initial_conditions}.  Secondly, those 20
systems have massive disks roughly comparable to the central protostar
mass as shown in Figure \ref{fig:disk_mass_all}.  This is consistent
with \citet{2010MNRAS.401.1505B}, but not with
\citet{2015MNRAS.450.3306F}.  The misaligned systems in
\citet{2015MNRAS.450.3306F} preferentially have less massive disks,
which are likely disturbed by the subsequent accretion from the
envelope and/or by the perturbation from a distant star.  The single
star systems in our simulation neglects the possible interaction with
the outer system, and may underestimate the possible evolution toward
the star-disk misalignment.  \citet{2012Natur.491..418B} showed that
the gravitational torque due to a distant star significantly affects
the orientation of the disk plane relative to the central stellar
spin.  Finally the sink particle technique is admittedly very
approximate and cannot reliably describe the physics inside the
accretion radius of the sink particle.

Having said so, however, it is encouraging that our higher-resolution
SPH simulations are generally consistent with the previous SPH result
by \citet{2010MNRAS.401.1505B}. Furthermore,
\citet{2015MNRAS.450.3306F} also found a star-disk alignment if the
disk mass is comparable to that of the protostar even in their AMR
simulation. Thus the star-disk mass ratio may be an important
parameter that is responsible for the degree of the primordial
star-disk orientation.

%%%%%%%%%%%%%%%%%%%%%%%%%%%%%%%%%%%%%%%%%%%%%%%%%%%%%%%%%%%%%%%%%%

%%% 5 Model Warped disk
\section{Warped Disk and Envelope Rotation Structures}
\label{sec:other_implications}
%%%%%%%%%%%%%%%%%%%%%%%%%%%%%%%%%%%%%%%%%%%%%%%%%%%%%%%%%%%%%%%%%%

In following two subsection, we examine whether our simulation results
can explain recent observations of the warped disk and
counter-rotating envelope.

%%% 5.1.
\subsection{Warped Disks}
\label{subsec:warped_disk}
%%%%%%%%%%%%%%%%%%%%%%%%%%%%%%%%%%%%%%%%%%%%%%%%%%%%%%%%%%%%%%%%%%

\citet{2019Natur.565..206S} reported the warped disk-like structure
around a young protostar, IRAS 04368+2557, located in the protostellar
core L1527 that is classified as a Class 0 YSO.  Because such a warped
disk is expected to evolve into spin-orbit misaligned planetary
systems, their detailed structure may be connected to the observed
diversity of the spin-orbit architecture.

\begin{figure*}%%% Figure 11
  \centering
  \includegraphics[clip,width=100mm,angle=-90]{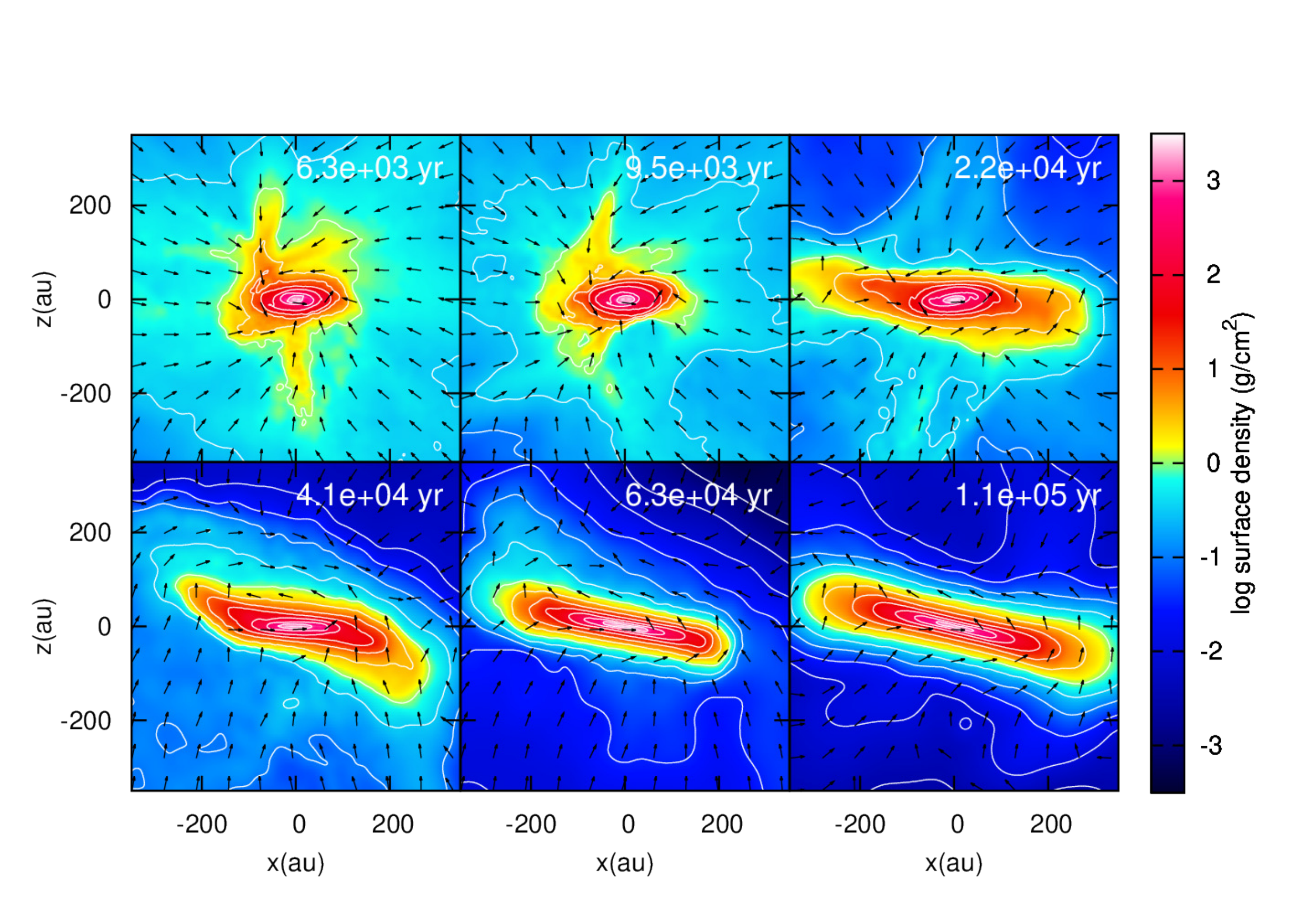}
  \caption{ The surface density evolution on the x-z plane for model
    \WarpModel.  The top-left, top-middle and top-right panels show
    snapshots at $6.3\times10^3$ yr, $9.5\times10^3$ yr and
    $2.2\times10^4$ yr after the protostar formation, respectively.
    The bottom-left, bottom-middle and bottom-right panels show
    snapshots at $4.1\times10^4$ yr, $6.3\times10^4$ yr and
    $1.1\times10^5$ yr after the protostar formation, respectively.
    White lines show contours of the surface density.  Black arrows
    show direction of the density weighted velocity.
    \label{fig:snapshots_xz_of_warp_model}
  }
\end{figure*}

\begin{figure*}%%% Figure 12
  \centering
  \includegraphics[clip,width=100mm,angle=-90]{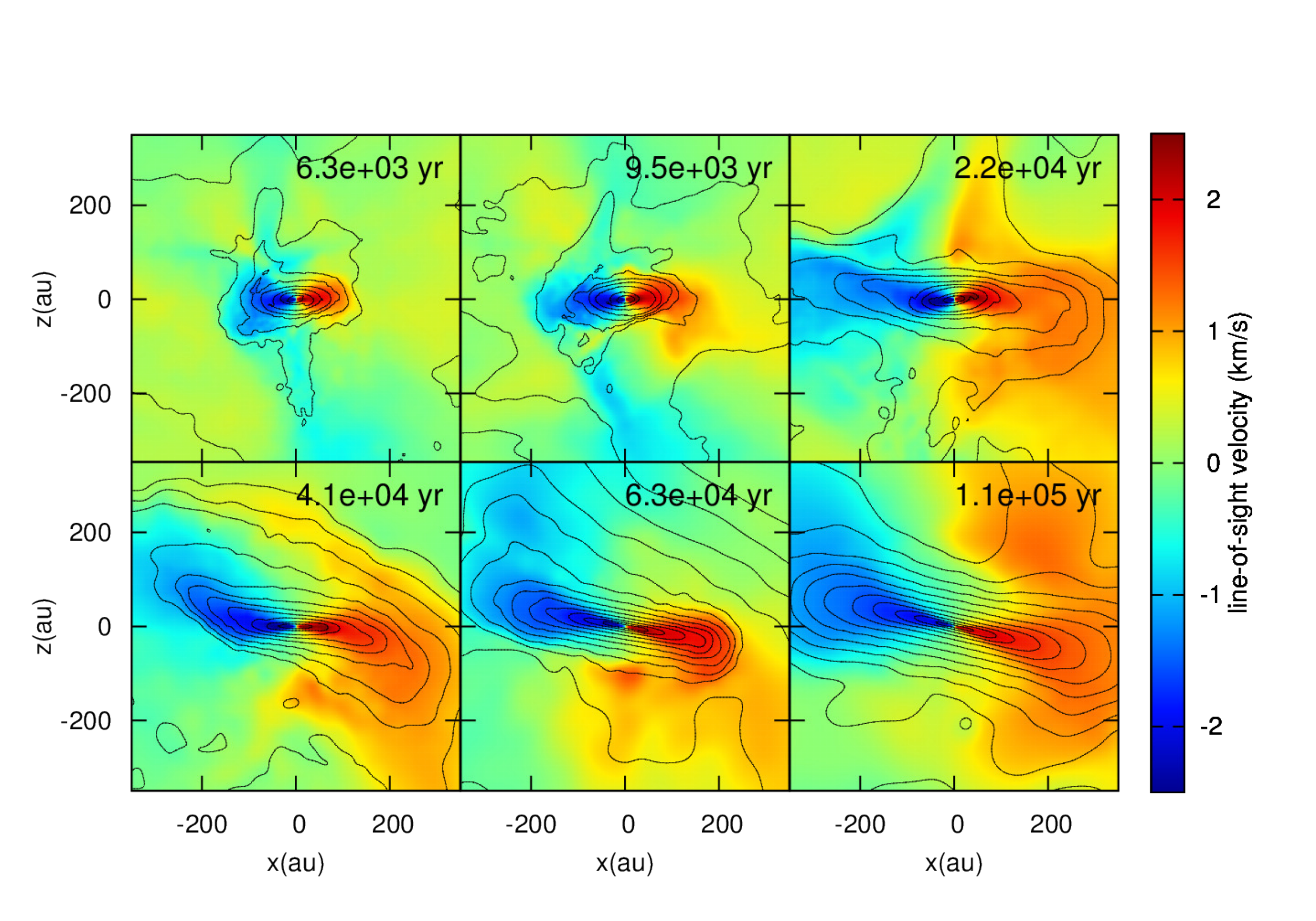}
  \caption{ The evolution of the density weighted line-of-sight
    velocity on the x-z plane for model \WarpModel.  The epochs of
    each panel are the same as Figure
    \ref{fig:snapshots_xz_of_warp_model}.  Black lines are contours of
    the surface density.
    \label{fig:Moment1_0303}
  }
\end{figure*}

We suggest that such a warped disk can be explained by the turbulence
in the molecular cloud cores.  We show the evolution of the surface
density and line-of-sight velocity of model \WarpModel in Figures
\ref{fig:snapshots_xz_of_warp_model} and \ref{fig:Moment1_0303}.
Figure \ref{fig:snapshots_xz_of_warp_model} shows that the warped
structure is formed in $t\lesssim 4\times 10^4$ yr.  In particular,
the top right and bottom left panels show the elongation of the
surface density and rotation structure.  The turbulent accretion flow
from the circumstellar envelope causes this warped disk structure.

%%%%%%%%%%%%%%%%%%%%%%%%%%%%%%
%%%%%%%%%%%%%%%%%%%%%%%%%%%%%%

%%%%%%%%%%%%%%%%%%%%%%%%%%%%%%
%%%%%%%%%%%%%%%%%%%%%%%%%%%%%%

%%%%%%%%%%%%%%%%%%%%%%%%%%%%%%
\begin{figure}%%% Figure 13
  \centering
  \includegraphics[clip,width=60mm,angle=-90]{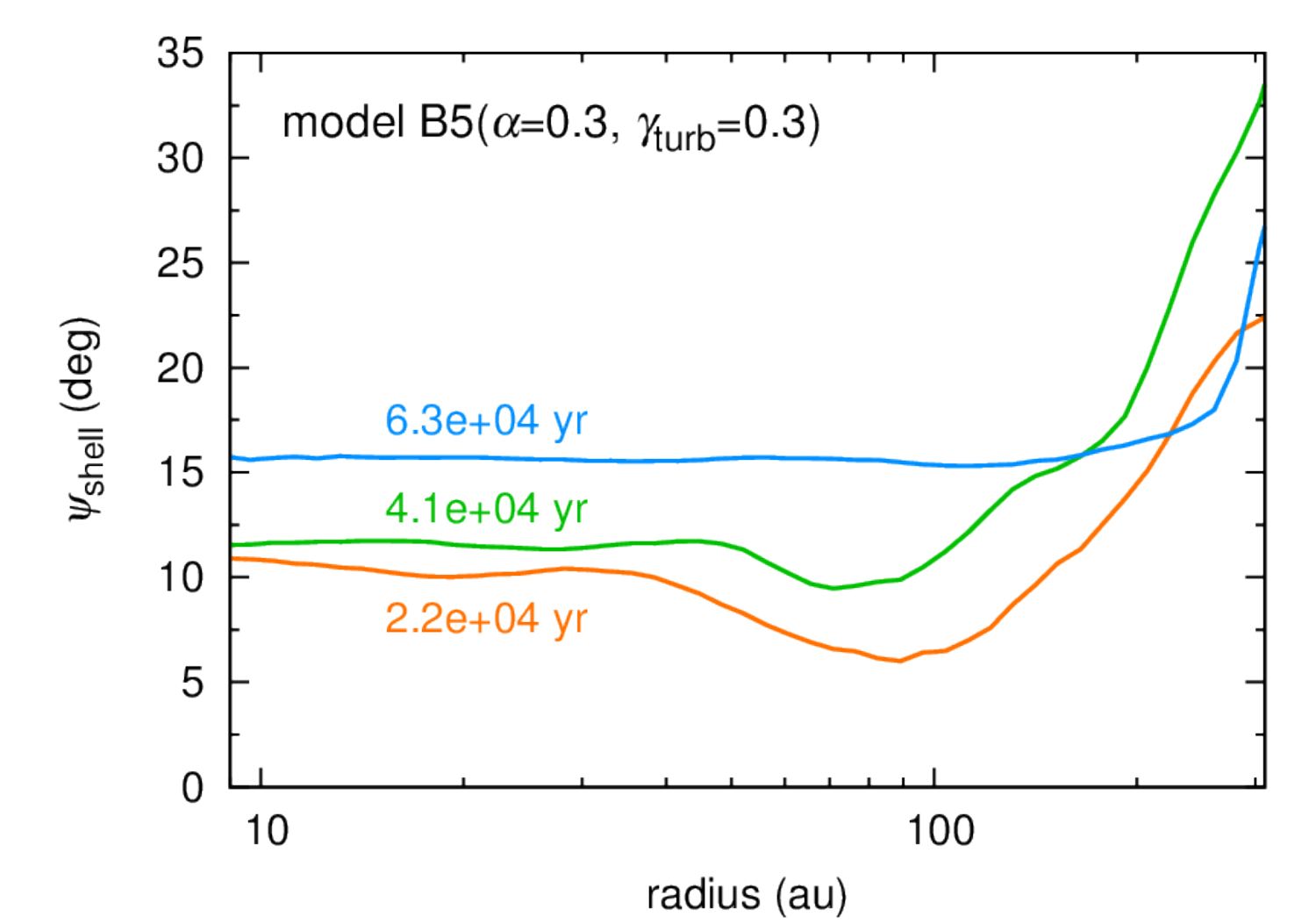}
  \caption{ Radial distribution of $\psi_{\rm shell}$ for model
    \WarpModel.  Orange, green, and blue lines correspond to
    $\psi_{\rm shell}$ at $2.2\times10^4$ yr, $4.1\times10^4$ yr, and
    $6.3\times10^4$ yr after the protostar formation, respectively.
    \label{fig:rp_warped}
  }
\end{figure}
%%%%%%%%%%%%%%%%%%%%%%%%%%%%%%

To examine the warped structure of the protoplanetary disk quantitatively,
we plot the angle $\pshell(r)$ between the angular momentum of the spherical shell
and the spin of the protostar,
%%%%%%%%%%%%%%%%%%%%%%%%%%%%%%
\begin{align}
  \label{eq:pshell}
  \pshell(r)
  = \cos^{-1} \left( \frac{\Js \cdot \Jshell(r)}{|\Js| |\Jshell(r)|} \right),
\end{align}
%%%%%%%%%%%%%%%%%%%%%%%%%%%%%%
where $\Jshell(r)$ is the angular momentum of the spherical shell at
$r$.

Figure \ref{fig:rp_warped} shows the radial profile of $\pshell(r)$ at
different epochs.  In all epochs, $\pshell(r)$ is almost flat in
$r<40$ au.  This means that the inner disk in $r<40$ au is not warped.
On the other hand, $\pshell(r)$ decreases in
$40~\rm{au}<r<100~\rm{au}$ at $t=2.2\times 10^4$ yr (orange) and at
$t=4.1\times 10^4$ yr (green) indicating that the disk is warped in
this region.  The relative angle between the inner and outer region is
$\sim5^{\circ}$ at $t=2.2\times 10^4$ yr and good agreement with
\citet{2019Natur.565..206S}.  This indicates that, with the turbulent
infalling envelope, the rotation axis of the inner disk is not
necessarily aligned with that of the outer disk and the warped disk is
expected in the early evolution phase of YSOs.

%%%%%%%%%%%%%%%%%%%%%%%%%%%%%%
\begin{figure}%%% Figure 14
  \centering
  \includegraphics[clip,width=60mm,angle=-90]{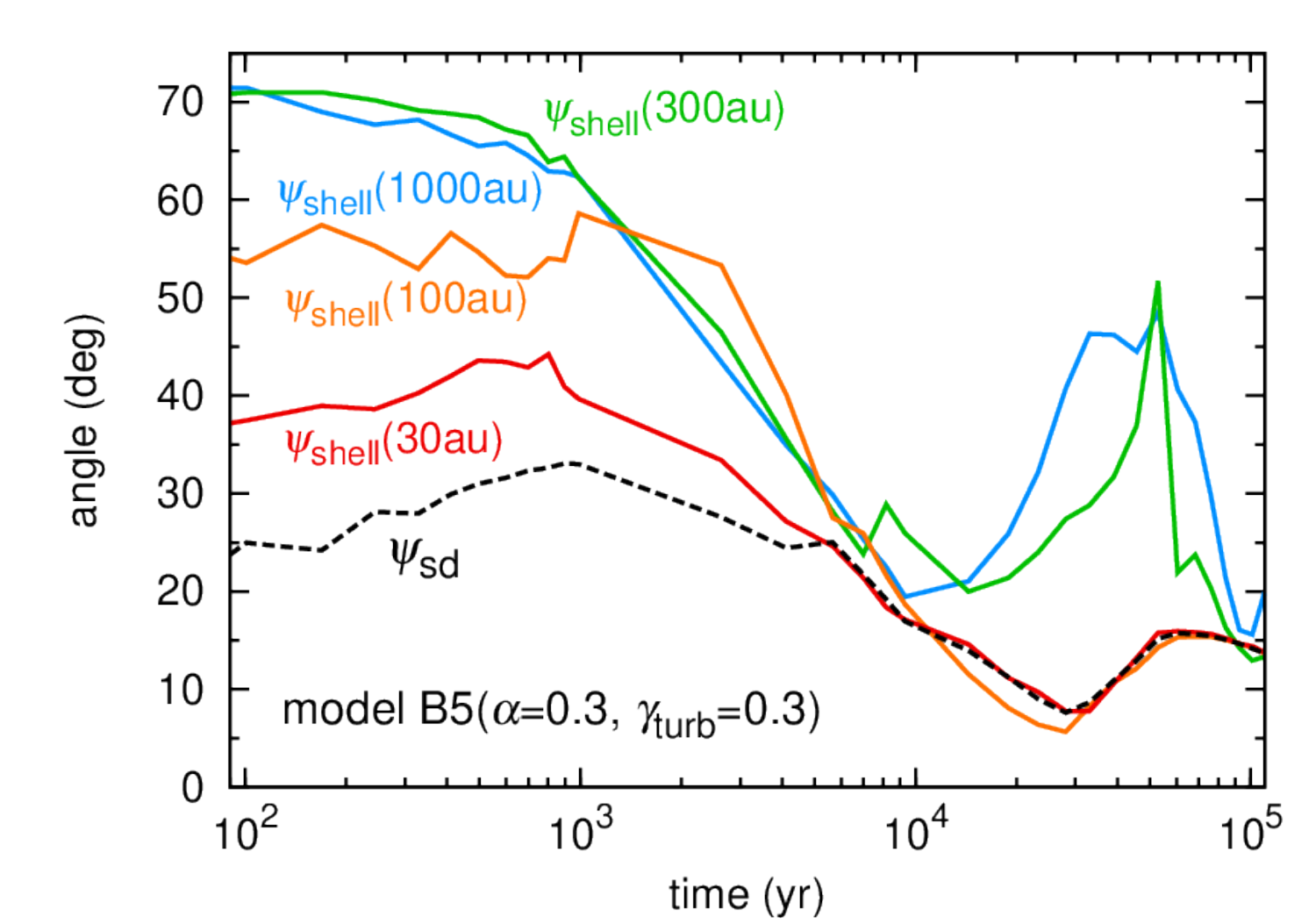}
  \caption{ Time evolution of $\psd$ and $\pshell$ for model
    \WarpModel.  Red, orange, green, and blue lines correspond to the
    $\pshell$ at $r=30$ au, $100$ au, $300$ au, and $1000$ au,
    respectively.  Dashed black line shows $\psd$.
    \label{fig:time_shell_warped}
  }
\end{figure}
%%%%%%%%%%%%%%%%%%%%%%%%%%%%%%

%%%%%%%%%%%%%%%%%%%%%%%%%%%%%%
\begin{figure}%%% Figure 15
  \centering
  \includegraphics[clip,width=60mm,angle=-90]{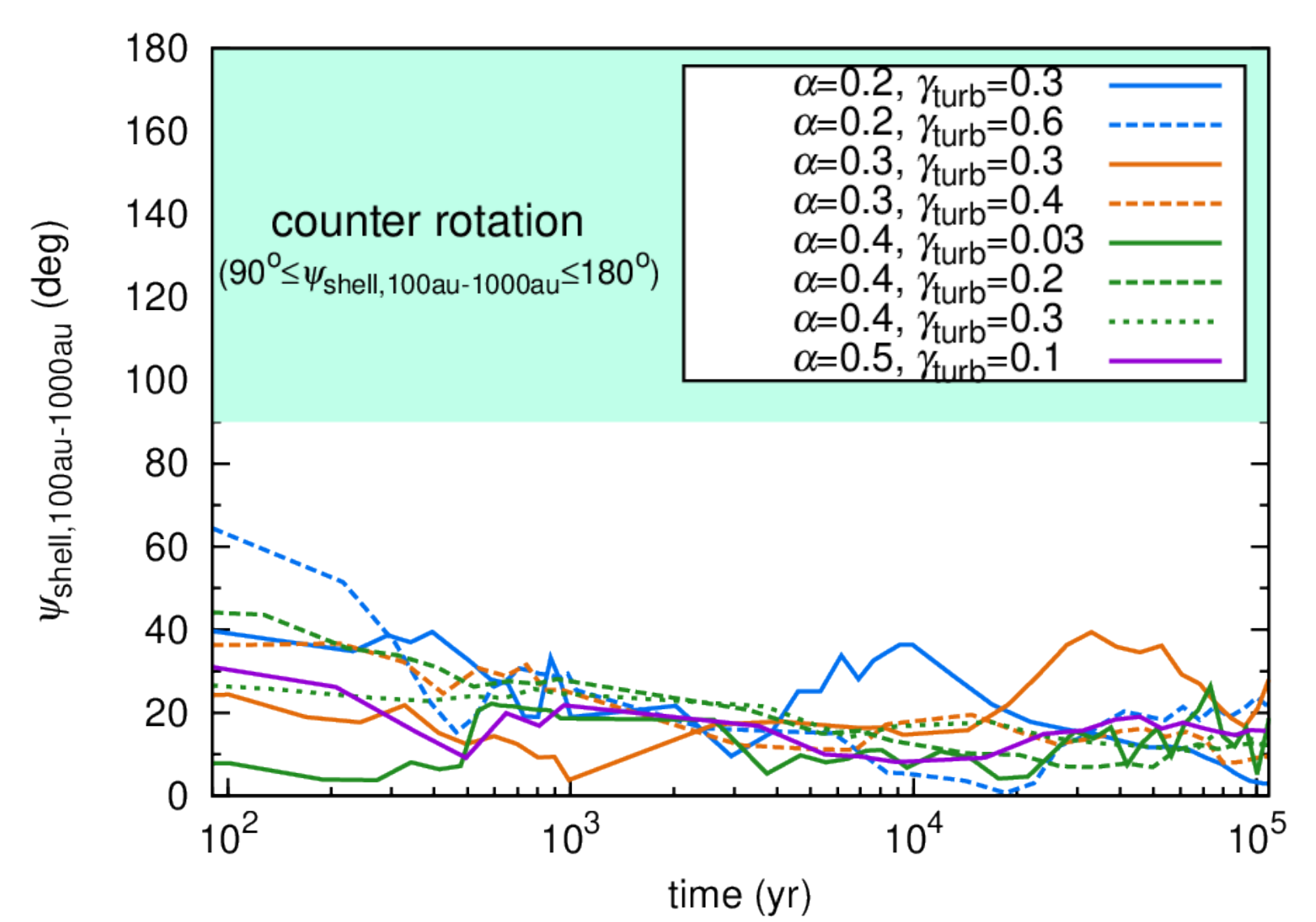}
  \caption{ Time evolution of $\psi_{{\rm{shell,100~au-1000~au}}}$ for
    all the simulation models with $\psd > 20^\circ$ at $t\sim10^2$
    yr.  The shaded region corresponds to a counter-rotating structure
    ($90^{\circ} \leq \psi_{{\rm{shell,100~au-1000~au}}} \leq
    180^{\circ}$).
    \label{fig:time_shell_in-out}
  }
\end{figure}
%%%%%%%%%%%%%%%%%%%%%%%%%%%%%%

Note that the top-left and top-middle panels of Figure
\ref{fig:snapshots_xz_of_warp_model} show the filamentary structure of
the infalling envelope which extends to the z directions.
Interestingly, the density weighted line-of-sight velocity of these
filaments are both blue (top-left and top-middle of Figure
\ref{fig:Moment1_0303}), meaning that the accretion flow has the same
direction in the upper and lower regions of the disk.

\citet{2014ApJ...793....1Y} reported the infalling flows of the
envelope in parabolic trajectories toward the Keplerian disk of a
Class I protostar, L1489 IRS.  The red-shifted and blue-shifted
structures in the lower-left and lower-center panels of Figure
\ref{fig:Moment1_0303} look very similar to the infalling envelope
structure reported in their Figure 3.  This suggests that the arc-like
structure of infalling envelopes may be naturally formed by the
turbulent accretion of the infalling matter in the early phase of
YSOs.

%%%%%%%%%%%%%%%%%%%%%%%%%%%%%%%%%%%%%%%%%%%%%%%%%%%%%%%%%%%%%%%%%%

%%% 5.2.
\subsection{Envelope rotation structure}
\label{counter-rotation}
%%%%%%%%%%%%%%%%%%%%%%%%%%%%%%%%%%%%%%%%%%%%%%%%%%%%%%%%%%%%%%%%%%

\citet{2018ApJ...865...51T} found a Class I YSO in which the rotation
direction of the circumstellar envelope significantly change from 1000
au scale to inner 100 au scale, which can be interpreted as a counter
rotation between the protoplanetary disk and circumstellar envelope.
The physical mechanism which induces such a counter-rotating structure
is still unclear.  One may expect that the random motion of the
turbulence may create the random rotation direction of the
circumstellar envelope, leading to a counter rotation.  However, we do
not find such a significant change of the rotation direction in the
circumstellar envelope in our simulations.  Rather, the protoplanetary
disk rotation tends to be aligned with the circumstellar envelope
rotation especially in the late phase.

An example is presented in Figure \ref{fig:time_shell_warped} that
shows the time evolution of $\psd$ and $\pshell$ at $r=30$ au (red),
$100$ au (orange), $300$ au (green) and $1000$ au (blue),
respectively, for model \WarpModel.  Even at $t=10^2$ yr, $\pshell$ at
$r=1000$ au is $\sim 70^{\circ}$ and $\pshell$ is not counter
rotating.  Subsequently, $\pshell$ keeps decreasing, instead of
increasing, and all the values of $\pshell$ as well as $\psd$ converge
to $\sim 15^{\circ}$.  This clearly indicates that the turbulence in
molecular cloud cores is unlikely to produce a counter-rotating
structure.

The prograde rotation inside an isolated compact region is a generic
outcome of the gravitational collapse of turbulent molecular cloud
cores.  In order to see it, we introduce the relative angle between
the angular momenta of the inner shell at $r=100$ au and the outer
shell at $r=1000$ au:
%%%%%%%%%%%%%%%%%%%%%%%%%%%%%%
\begin{align}
  \label{eq:pshell_in-out}
  \psi_{{\rm{shell,100~au-1000~au}}} = \cos^{-1}
  \left(
  \frac{\Jshell({\rm{100~au}}) \cdot \Jshell({\rm{1000~au}})}
       {|\Jshell({\rm{100~au}})||\Jshell({\rm{1000~au}})|}
       \right).
\end{align}
%%%%%%%%%%%%%%%%%%%%%%%%%%%%%%

Figure \ref{fig:time_shell_in-out} shows
$\psi_{{\rm{shell,100~au-1000~au}}}$ of all the simulation models with
star-disk misalignment ($\psd > 20^{\circ}$ at $t\sim10^2$ yr).
Figure \ref{fig:time_shell_in-out} suggests that
$\psi_{{\rm{shell,100~au-1000~au}}}$ is $\lesssim 70^{\circ}$ even at
$t=10^2$ yr, and then gradually becomes aligned towards $\lesssim
20^{\circ}$ at $t=10^5$ yr.  Thus, no simulation exhibits the
misalignment between the inner envelope ($r \sim 100$ au) and the
outer envelope ($r \sim 1000$ au).  We compared the angular momentum
of the inner shell ($r=100$ au) with that of the further outer shells
($1000~\rm{au}<r<3000~\rm{au}$), and confirmed that the
counter-rotating envelope does not appear even in the scale of $3000$
au.

Thus we conclude that the turbulence in the molecular cloud core may
not create a counter-rotating envelope. Rather, the magnetic field in
the molecular cloud core may create it
\citep[e.g.,][]{2011ApJ...733...54K, 2011ApJ...738..180L,
  2015ApJ...810L..26T, 2016MNRAS.457.1037W, 2017PASJ...69...95T,
  2017MNRAS.466.1788W, 2018MNRAS.475.1859W, 2019MNRAS.tmp..933W}.

%%%%%%%%%%%%%%%%%%%%%%%%%%%%%%%%%%%%%%%%%%%%%%%%%%%%%%%%%%%%%%%%%%

% 6. conclusion
\section{Conclusion}
\label{sec:conclusion}
%%%%%%%%%%%%%%%%%%%%%%%%%%%%%%%%%%%%%%%%%%%%%%%%%%%%%%%%%%%%%

Observed exoplanetary systems are known to exhibit diverse properties
that are quite different from those of our Solar system. In
particular, the presence of the spin-orbit misaligned planetary
systems is supposed to carry important information concerning the
initial condition of the protoplanetary disk and the subsequent
formation and dynamical evolution of multi-planetary systems.

One of the basic questions underlying the spin-orbit architecture is
to what extent the spin axis of the protostar and the rotation axis of
the protoplanetary disk are aligned. While this question seems
well-defined and straightforward, it is not easy to give an
unambiguous answer because a variety of complicated physical processes
of very different spatial and time scales are involved.  Indeed, a
pioneering work by \citet{2010MNRAS.401.1505B} indicates that the
star-disk angle of the protoplanetary disk systems out of supersonic
turbulent clouds can be significantly misaligned, but that the
reliable prediction is not easy because the process occurs in an
inherently chaotic environment.

We have performed the SPH simulation of the collapse of turbulent
molecular cloud cores with varying the thermal and turbulent energy
contributions relative to the gravitational energy of those systems.
This paper has focused on the analysis of 20 single star-forming
systems out of the 26 models in total.  Our major findings are
summarized as follows.

\begin{description}
\item[1.] At the initial phase of the protostar formation, the axis of
  the stellar spin is not necessarily aligned with that of the disk
  rotation. The star-disk angle $\psd$ is almost randomly distributed
  within $\sim 130^{\circ}$ until $\sim 10^{4}$ yr after the protostar
  formation.
\item[2.] The subsequent mass accretion from the disk to the protostar
  gradually aligns the stellar spin toward the disk rotation axis.
  The disk also receives the angular momentum accretion from the
  surrounding envelope, and its rotation axis becomes aligned to that
  of the initial angular momentum of the cloud core. As a result,
  $\psd$ becomes less than $\sim 20^{\circ}$ in $\sim 10^{4}$ yr after
  the protostar formation. The timescale of the star-disk alignment,
  $t_{\rm alignment} \sim 10^{4}$ yr corresponds to a typical mass
  doubling time of the central protostar.
\item[3.] The star-disk angles $\psd$, measured at the epoch of the
  protostar formation (about $t=10^2$ yr) and the end of our
  simulations ($t=10^5$ yr) are insensitive to $\alpha=E_{\rm
    thermal}/|E_{\rm gravity}|$ nor to $\gturb=E_{\rm
    turbulence}/|E_{\rm gravity}|$.
\item[4.] Our simulation sometimes produces a warped disk structure as
  recently reported by \citet{2019Natur.565..206S}. A clear warped
  structure is produced when the mass accretion and angular momentum
  transfer from the envelope to the outer disk proceeds along the
  direction significantly different from that of the existing inner
  disk. This process also changes the rotation axis of the inner disk
  gradually, and $\psd$ fluctuates by an amount of $\sim 10^{\circ}$
  even after it once becomes less than $20^{\circ}$.
\item[5.] Rotation directions of the disk and envelope are generally
  well aligned, especially after the significant mass accretion ceases
  ($t \sim 10^5$ yr).  Therefore the turbulence of the molecular cloud
  cores alone does not lead to a counter-rotating disk structure.
\end{description}

Our overall conclusion is that the stellar spin and disk rotation axes
of a protoplanetary disk system out of a turbulent cloud core are
aligned less than $\sim 20^{\circ}$. We should emphasize, however,
that this conclusion holds only for an isolated single star-forming
case.  If the initial cloud core has sufficiently massive and its
thermal and turbulent energies are smaller than the gravitational
energy, it would preferentially produce multiple protoplanetary disks
inside (see Figure \ref{fig:alpha-gamma}). Then the star-disk angle of
a planetary system can be significantly affected by the perturbation
from a nearby system as proposed by \citet{2012Natur.491..418B}, for
instance. Furthermore, the magnetic field, which is neglected in the
present simulation, may also play an important role. More realistic
simulations including the magnetic field and turbulence simultaneously
are numerically demanding and expensive, but we plan to perform and
hope to report the result in a future work.

\section*{Acknowledgements}

We thank an anonymous referee for a number of important and
constructive comments that significantly improved the earlier
manuscript of the paper.  Numerical computations were in part
carried out on Cray XC50 at Center for Computational Astrophysics,
National Astronomical Observatory of Japan.  This research is
supported by JSPS (Japan Society of Promotion of Science) Core-to-Core
Program ``International Network of Planetary Sciences'', by the
Astrobiology Center of National Institutes of Natural Sciences (NINS)
Grant Number AB311025, and also by JSPS KAKENHI Grant Numbers 18H01247
(Y.S.), 18H05437 (Y.T.), 18K13581 (Y.T.), and 19H01947 (Y.S.).

%%%%%%%%%%%%%%%%%%%%%%%%%%%%%%%%%%%%%%%%%%%%%%%%%%


\bibliographystyle{mnras}
%\bibliography{article_takaishids}

\begin{thebibliography}{}
\makeatletter
\relax
\def\mn@urlcharsother{\let\do\@makeother \do\$\do\&\do\#\do\^\do\_\do\%\do\~}
\def\mn@doi{\begingroup\mn@urlcharsother \@ifnextchar [ {\mn@doi@}
  {\mn@doi@[]}}
\def\mn@doi@[#1]#2{\def\@tempa{#1}\ifx\@tempa\@empty \href
  {http://dx.doi.org/#2} {doi:#2}\else \href {http://dx.doi.org/#2} {#1}\fi
  \endgroup}
\def\mn@eprint#1#2{\mn@eprint@#1:#2::\@nil}
\def\mn@eprint@arXiv#1{\href {http://arxiv.org/abs/#1} {{\tt arXiv:#1}}}
\def\mn@eprint@dblp#1{\href {http://dblp.uni-trier.de/rec/bibtex/#1.xml}
  {dblp:#1}}
\def\mn@eprint@#1:#2:#3:#4\@nil{\def\@tempa {#1}\def\@tempb {#2}\def\@tempc
  {#3}\ifx \@tempc \@empty \let \@tempc \@tempb \let \@tempb \@tempa \fi \ifx
  \@tempb \@empty \def\@tempb {arXiv}\fi \@ifundefined
  {mn@eprint@\@tempb}{\@tempb:\@tempc}{\expandafter \expandafter \csname
  mn@eprint@\@tempb\endcsname \expandafter{\@tempc}}}

\bibitem[\protect\citeauthoryear{{Albrecht} et~al.,}{{Albrecht}
  et~al.}{2012}]{2012ApJ...757...18A}
{Albrecht} S.,  et~al., 2012, \mn@doi [\apj] {10.1088/0004-637X/757/1/18},
  \href {https://ui.adsabs.harvard.edu/\#abs/2012ApJ...757...18A} {757, 18}

\bibitem[\protect\citeauthoryear{{Albrecht}, {Winn}, {Marcy}, {Howard},
  {Isaacson}  \& {Johnson}}{{Albrecht} et~al.}{2013}]{2013ApJ...771...11A}
{Albrecht} S.,  {Winn} J.~N.,  {Marcy} G.~W.,  {Howard} A.~W.,  {Isaacson} H.,
   {Johnson} J.~A.,  2013, \mn@doi [\apj] {10.1088/0004-637X/771/1/11}, \href
  {https://ui.adsabs.harvard.edu/abs/2013ApJ...771...11A} {771, 11}

\bibitem[\protect\citeauthoryear{{Alibert}, {Mordasini}, {Benz}  \&
  {Winisdoerffer}}{{Alibert} et~al.}{2005}]{2005A&A...434..343A}
{Alibert} Y.,  {Mordasini} C.,  {Benz} W.,   {Winisdoerffer} C.,  2005, \mn@doi
  [\aap] {10.1051/0004-6361:20042032}, \href
  {https://ui.adsabs.harvard.edu/abs/2005A&A...434..343A} {434, 343}

\bibitem[\protect\citeauthoryear{{Anderson}, {Storch}  \& {Lai}}{{Anderson}
  et~al.}{2016}]{2016MNRAS.456.3671A}
{Anderson} K.~R.,  {Storch} N.~I.,   {Lai} D.,  2016, \mn@doi [\mnras]
  {10.1093/mnras/stv2906}, \href
  {https://ui.adsabs.harvard.edu/abs/2016MNRAS.456.3671A} {456, 3671}

\bibitem[\protect\citeauthoryear{{Andre}, {Ward-Thompson}  \& {Motte}}{{Andre}
  et~al.}{1996}]{1996A&A...314..625A}
{Andre} P.,  {Ward-Thompson} D.,   {Motte} F.,  1996, \aap, \href
  {https://ui.adsabs.harvard.edu/abs/1996A&A...314..625A} {314, 625}

\bibitem[\protect\citeauthoryear{{Barranco} \& {Goodman}}{{Barranco} \&
  {Goodman}}{1998}]{1998ApJ...504..207B}
{Barranco} J.~A.,  {Goodman} A.~A.,  1998, \mn@doi [\apj] {10.1086/306044},
  \href {http://ads.nao.ac.jp/abs/1998ApJ...504..207B} {504, 207}

\bibitem[\protect\citeauthoryear{{Bate}}{{Bate}}{2018}]{2018MNRAS.475.5618B}
{Bate} M.~R.,  2018, \mn@doi [\mnras] {10.1093/mnras/sty169}, \href
  {https://ui.adsabs.harvard.edu/abs/2018MNRAS.475.5618B} {475, 5618}

\bibitem[\protect\citeauthoryear{{Bate} \& {Burkert}}{{Bate} \&
  {Burkert}}{1997}]{1997MNRAS.288.1060B}
{Bate} M.~R.,  {Burkert} A.,  1997, \mnras, \href
  {http://adsabs.harvard.edu/abs/1997MNRAS.288.1060B} {288, 1060}

\bibitem[\protect\citeauthoryear{{Bate}, {Bonnell}  \& {Price}}{{Bate}
  et~al.}{1995}]{1995MNRAS.277..362B}
{Bate} M.~R.,  {Bonnell} I.~A.,   {Price} N.~M.,  1995, \mnras, \href
  {http://ads.nao.ac.jp/abs/1995MNRAS.277..362B} {277, 362}

\bibitem[\protect\citeauthoryear{{Bate}, {Lodato}  \& {Pringle}}{{Bate}
  et~al.}{2010}]{2010MNRAS.401.1505B}
{Bate} M.~R.,  {Lodato} G.,   {Pringle} J.~E.,  2010, \mn@doi [\mnras]
  {10.1111/j.1365-2966.2009.15773.x}, \href
  {https://ui.adsabs.harvard.edu/\#abs/2010MNRAS.401.1505B} {401, 1505}

\bibitem[\protect\citeauthoryear{{Batygin}}{{Batygin}}{2012}]{2012Natur.491..418B}
{Batygin} K.,  2012, \mn@doi [\nat] {10.1038/nature11560}, \href
  {https://ui.adsabs.harvard.edu/abs/2012Natur.491..418B} {491, 418}

\bibitem[\protect\citeauthoryear{{Beaug{\'e}} \& {Nesvorn{\'y}}}{{Beaug{\'e}}
  \& {Nesvorn{\'y}}}{2012}]{2012ApJ...751..119B}
{Beaug{\'e}} C.,  {Nesvorn{\'y}} D.,  2012, \mn@doi [\apj]
  {10.1088/0004-637X/751/2/119}, \href
  {https://ui.adsabs.harvard.edu/abs/2012ApJ...751..119B} {751, 119}

\bibitem[\protect\citeauthoryear{{Bertoldi} \& {McKee}}{{Bertoldi} \&
  {McKee}}{1992}]{1992ApJ...395..140B}
{Bertoldi} F.,  {McKee} C.~F.,  1992, \mn@doi [\apj] {10.1086/171638}, \href
  {https://ui.adsabs.harvard.edu/abs/1992ApJ...395..140B} {395, 140}

\bibitem[\protect\citeauthoryear{{Burkert} \& {Bodenheimer}}{{Burkert} \&
  {Bodenheimer}}{2000}]{2000ApJ...543..822B}
{Burkert} A.,  {Bodenheimer} P.,  2000, \mn@doi [\apj] {10.1086/317122}, \href
  {http://ads.nao.ac.jp/abs/2000ApJ...543..822B} {543, 822}

\bibitem[\protect\citeauthoryear{{Butler} \& {Tan}}{{Butler} \&
  {Tan}}{2012}]{2012ApJ...754....5B}
{Butler} M.~J.,  {Tan} J.~C.,  2012, \mn@doi [\apj]
  {10.1088/0004-637X/754/1/5}, \href
  {https://ui.adsabs.harvard.edu/abs/2012ApJ...754....5B} {754, 5}

\bibitem[\protect\citeauthoryear{{Crutcher}, {Nutter}, {Ward-Thompson}  \&
  {Kirk}}{{Crutcher} et~al.}{2004}]{2004ApJ...600..279C}
{Crutcher} R.~M.,  {Nutter} D.~J.,  {Ward-Thompson} D.,   {Kirk} J.~M.,  2004,
  \mn@doi [\apj] {10.1086/379705}, \href
  {https://ui.adsabs.harvard.edu/abs/2004ApJ...600..279C} {600, 279}

\bibitem[\protect\citeauthoryear{{Fabrycky} \& {Tremaine}}{{Fabrycky} \&
  {Tremaine}}{2007}]{2007ApJ...669.1298F}
{Fabrycky} D.,  {Tremaine} S.,  2007, \mn@doi [\apj] {10.1086/521702}, \href
  {https://ui.adsabs.harvard.edu/abs/2007ApJ...669.1298F} {669, 1298}

\bibitem[\protect\citeauthoryear{{Fielding}, {McKee}, {Socrates}, {Cunningham}
  \& {Klein}}{{Fielding} et~al.}{2015}]{2015MNRAS.450.3306F}
{Fielding} D.~B.,  {McKee} C.~F.,  {Socrates} A.,  {Cunningham} A.~J.,
  {Klein} R.~I.,  2015, \mn@doi [\mnras] {10.1093/mnras/stv836}, \href
  {https://ui.adsabs.harvard.edu/abs/2015MNRAS.450.3306F} {450, 3306}

\bibitem[\protect\citeauthoryear{{Gingold} \& {Monaghan}}{{Gingold} \&
  {Monaghan}}{1977}]{1977MNRAS.181..375G}
{Gingold} R.~A.,  {Monaghan} J.~J.,  1977, \mn@doi [\mnras]
  {10.1093/mnras/181.3.375}, \href
  {http://adsabs.harvard.edu/abs/1977MNRAS.181..375G} {181, 375}

\bibitem[\protect\citeauthoryear{{Goodman}, {Benson}, {Fuller}  \&
  {Myers}}{{Goodman} et~al.}{1993}]{1993ApJ...406..528G}
{Goodman} A.~A.,  {Benson} P.~J.,  {Fuller} G.~A.,   {Myers} P.~C.,  1993,
  \mn@doi [\apj] {10.1086/172465}, \href
  {http://ads.nao.ac.jp/abs/1993ApJ...406..528G} {406, 528}

\bibitem[\protect\citeauthoryear{{Hillenbrand}}{{Hillenbrand}}{1997}]{1997AJ....113.1733H}
{Hillenbrand} L.~A.,  1997, \mn@doi [\aj] {10.1086/118389}, \href
  {https://ui.adsabs.harvard.edu/abs/1997AJ....113.1733H} {113, 1733}

\bibitem[\protect\citeauthoryear{{Hirano}, {Suto}, {Winn}, {Taruya}, {Narita},
  {Albrecht}  \& {Sato}}{{Hirano} et~al.}{2011}]{2011ApJ...742...69H}
{Hirano} T.,  {Suto} Y.,  {Winn} J.~N.,  {Taruya} A.,  {Narita} N.,  {Albrecht}
  S.,   {Sato} B.,  2011, \mn@doi [\apj] {10.1088/0004-637X/742/2/69}, \href
  {https://ui.adsabs.harvard.edu/abs/2011ApJ...742...69H} {742, 69}

\bibitem[\protect\citeauthoryear{{Hirano} et~al.,}{{Hirano}
  et~al.}{2012}]{2012ApJ...759L..36H}
{Hirano} T.,  et~al., 2012, \mn@doi [\apjl] {10.1088/2041-8205/759/2/L36},
  \href {https://ui.adsabs.harvard.edu/abs/2012ApJ...759L..36H} {759, L36}

\bibitem[\protect\citeauthoryear{{Huber} et~al.,}{{Huber}
  et~al.}{2013}]{2013Sci...342..331H}
{Huber} D.,  et~al., 2013, \mn@doi [Science] {10.1126/science.1242066}, \href
  {https://ui.adsabs.harvard.edu/\#abs/2013Sci...342..331H} {342, 331}

\bibitem[\protect\citeauthoryear{{Inutsuka}}{{Inutsuka}}{2012}]{2012PTEP.2012aA307I}
{Inutsuka} S.,  2012, \mn@doi [Progress of Theoretical and Experimental
  Physics] {10.1093/ptep/pts024}, \href
  {https://ui.adsabs.harvard.edu/abs/2012PTEP.2012aA307I} {2012, 01A307}

\bibitem[\protect\citeauthoryear{{Kamiaka}, {Benomar}, {Suto}, {Dai}, {Masuda}
  \& {Winn}}{{Kamiaka} et~al.}{2019}]{2019AJ....157..137K}
{Kamiaka} S.,  {Benomar} O.,  {Suto} Y.,  {Dai} F.,  {Masuda} K.,   {Winn}
  J.~N.,  2019, \mn@doi [\aj] {10.3847/1538-3881/ab04a9}, \href
  {https://ui.adsabs.harvard.edu/abs/2019AJ....157..137K} {157, 137}

\bibitem[\protect\citeauthoryear{{Kozai}}{{Kozai}}{1962}]{1962AJ.....67..591K}
{Kozai} Y.,  1962, \mn@doi [\aj] {10.1086/108790}, \href
  {http://ads.nao.ac.jp/abs/1962AJ.....67..591K} {67, 591}

\bibitem[\protect\citeauthoryear{{Krasnopolsky}, {Li}  \&
  {Shang}}{{Krasnopolsky} et~al.}{2011}]{2011ApJ...733...54K}
{Krasnopolsky} R.,  {Li} Z.-Y.,   {Shang} H.,  2011, \mn@doi [\apj]
  {10.1088/0004-637X/733/1/54}, \href
  {http://adsabs.harvard.edu/abs/2011ApJ...733...54K} {733, 54}

\bibitem[\protect\citeauthoryear{{Li}, {Krasnopolsky}  \& {Shang}}{{Li}
  et~al.}{2011}]{2011ApJ...738..180L}
{Li} Z.-Y.,  {Krasnopolsky} R.,   {Shang} H.,  2011, \mn@doi [\apj]
  {10.1088/0004-637X/738/2/180}, \href
  {http://adsabs.harvard.edu/abs/2011ApJ...738..180L} {738, 180}

\bibitem[\protect\citeauthoryear{{Lidov}}{{Lidov}}{1962}]{1962P&SS....9..719L}
{Lidov} M.~L.,  1962, \mn@doi [\planss] {10.1016/0032-0633(62)90129-0}, \href
  {https://ui.adsabs.harvard.edu/abs/1962P&SS....9..719L} {9, 719}

\bibitem[\protect\citeauthoryear{{Lin}, {Bodenheimer}  \& {Richardson}}{{Lin}
  et~al.}{1996}]{1996Natur.380..606L}
{Lin} D.~N.~C.,  {Bodenheimer} P.,   {Richardson} D.~C.,  1996, \mn@doi [\nat]
  {10.1038/380606a0}, \href
  {https://ui.adsabs.harvard.edu/abs/1996Natur.380..606L} {380, 606}

\bibitem[\protect\citeauthoryear{{Lucy}}{{Lucy}}{1977}]{1977AJ.....82.1013L}
{Lucy} L.~B.,  1977, \mn@doi [\aj] {10.1086/112164}, \href
  {http://adsabs.harvard.edu/abs/1977AJ.....82.1013L} {82, 1013}

\bibitem[\protect\citeauthoryear{{Machida}, {Inutsuka}  \&
  {Matsumoto}}{{Machida} et~al.}{2007}]{2007ApJ...670.1198M}
{Machida} M.~N.,  {Inutsuka} S.,   {Matsumoto} T.,  2007, \mn@doi [\apj]
  {10.1086/521779}, \href {http://adsabs.harvard.edu/abs/2007ApJ...670.1198M}
  {670, 1198}

\bibitem[\protect\citeauthoryear{{Machida}, {Inutsuka}  \&
  {Matsumoto}}{{Machida} et~al.}{2010}]{2010ApJ...724.1006M}
{Machida} M.~N.,  {Inutsuka} S.,   {Matsumoto} T.,  2010, \mn@doi [\apj]
  {10.1088/0004-637X/724/2/1006}, \href
  {http://ads.nao.ac.jp/abs/2010ApJ...724.1006M} {724, 1006}

\bibitem[\protect\citeauthoryear{{Machida}, {Inutsuka}  \&
  {Matsumoto}}{{Machida} et~al.}{2014}]{2014MNRAS.438.2278M}
{Machida} M.~N.,  {Inutsuka} S.,   {Matsumoto} T.,  2014, \mn@doi [\mnras]
  {10.1093/mnras/stt2343}, \href
  {https://ui.adsabs.harvard.edu/abs/2014MNRAS.438.2278M} {438, 2278}

\bibitem[\protect\citeauthoryear{{Masunaga} \& {Inutsuka}}{{Masunaga} \&
  {Inutsuka}}{2000}]{2000ApJ...531..350M}
{Masunaga} H.,  {Inutsuka} S.,  2000, \mn@doi [\apj] {10.1086/308439}, \href
  {http://ads.nao.ac.jp/abs/2000ApJ...531..350M} {531, 350}

\bibitem[\protect\citeauthoryear{{McKee} \& {Ostriker}}{{McKee} \&
  {Ostriker}}{2007}]{2007ARA&A..45..565M}
{McKee} C.~F.,  {Ostriker} E.~C.,  2007, \mn@doi [\araa]
  {10.1146/annurev.astro.45.051806.110602}, \href
  {https://ui.adsabs.harvard.edu/abs/2007ARA&A..45..565M} {45, 565}

\bibitem[\protect\citeauthoryear{{McLaughlin}}{{McLaughlin}}{1924}]{1924ApJ....60...22M}
{McLaughlin} D.~B.,  1924, \mn@doi [\apj] {10.1086/142826}, \href
  {https://ui.adsabs.harvard.edu/\#abs/1924ApJ....60...22M} {60, 22}

\bibitem[\protect\citeauthoryear{{Miyama}, {Hayashi}  \& {Narita}}{{Miyama}
  et~al.}{1984}]{1984ApJ...279..621M}
{Miyama} S.~M.,  {Hayashi} C.,   {Narita} S.,  1984, \mn@doi [\apj]
  {10.1086/161926}, \href
  {https://ui.adsabs.harvard.edu/abs/1984ApJ...279..621M} {279, 621}

\bibitem[\protect\citeauthoryear{{Monaghan} \& {Lattanzio}}{{Monaghan} \&
  {Lattanzio}}{1985}]{1985A&A...149..135M}
{Monaghan} J.~J.,  {Lattanzio} J.~C.,  1985, \aap, \href
  {http://ads.nao.ac.jp/abs/1985A%26A...149..135M} {149, 135}

\bibitem[\protect\citeauthoryear{{Nagasawa} \& {Ida}}{{Nagasawa} \&
  {Ida}}{2011}]{2011ApJ...742...72N}
{Nagasawa} M.,  {Ida} S.,  2011, \mn@doi [\apj] {10.1088/0004-637X/742/2/72},
  \href {https://ui.adsabs.harvard.edu/\#abs/2011ApJ...742...72N} {742, 72}

\bibitem[\protect\citeauthoryear{{Nagasawa}, {Ida}  \& {Bessho}}{{Nagasawa}
  et~al.}{2008}]{2008ApJ...678..498N}
{Nagasawa} M.,  {Ida} S.,   {Bessho} T.,  2008, \mn@doi [\apj]
  {10.1086/529369}, \href
  {https://ui.adsabs.harvard.edu/\#abs/2008ApJ...678..498N} {678, 498}

\bibitem[\protect\citeauthoryear{{Ohta}, {Taruya}  \& {Suto}}{{Ohta}
  et~al.}{2005}]{2005ApJ...622.1118O}
{Ohta} Y.,  {Taruya} A.,   {Suto} Y.,  2005, \mn@doi [\apj] {10.1086/428344},
  \href {https://ui.adsabs.harvard.edu/\#abs/2005ApJ...622.1118O} {622, 1118}

\bibitem[\protect\citeauthoryear{{Queloz}, {Eggenberger}, {Mayor}, {Perrier},
  {Beuzit}, {Naef}, {Sivan}  \& {Udry}}{{Queloz}
  et~al.}{2000}]{2000A&A...359L..13Q}
{Queloz} D.,  {Eggenberger} A.,  {Mayor} M.,  {Perrier} C.,  {Beuzit} J.~L.,
  {Naef} D.,  {Sivan} J.~P.,   {Udry} S.,  2000, \aap, \href
  {https://ui.adsabs.harvard.edu/\#abs/2000A&A...359L..13Q} {359, L13}

\bibitem[\protect\citeauthoryear{{Rasio} \& {Ford}}{{Rasio} \&
  {Ford}}{1996}]{1996Sci...274..954R}
{Rasio} F.~A.,  {Ford} E.~B.,  1996, \mn@doi [Science]
  {10.1126/science.274.5289.954}, \href
  {https://ui.adsabs.harvard.edu/abs/1996Sci...274..954R} {274, 954}

\bibitem[\protect\citeauthoryear{{Rossiter}}{{Rossiter}}{1924}]{1924ApJ....60...15R}
{Rossiter} R.~A.,  1924, \mn@doi [\apj] {10.1086/142825}, \href
  {https://ui.adsabs.harvard.edu/\#abs/1924ApJ....60...15R} {60, 15}

\bibitem[\protect\citeauthoryear{{Sakai}, {Hanawa}, {Zhang}, {Higuchi},
  {Ohashi}, {Oya}  \& {Yamamoto}}{{Sakai} et~al.}{2019}]{2019Natur.565..206S}
{Sakai} N.,  {Hanawa} T.,  {Zhang} Y.,  {Higuchi} A.~E.,  {Ohashi} S.,  {Oya}
  Y.,   {Yamamoto} S.,  2019, \mn@doi [\nat] {10.1038/s41586-018-0819-2}, \href
  {https://ui.adsabs.harvard.edu/abs/2019Natur.565..206S} {565, 206}

\bibitem[\protect\citeauthoryear{{Shu}}{{Shu}}{1977}]{1977ApJ...214..488S}
{Shu} F.~H.,  1977, \mn@doi [\apj] {10.1086/155274}, \href
  {http://adsabs.harvard.edu/abs/1977ApJ...214..488S} {214, 488}

\bibitem[\protect\citeauthoryear{{Tafalla}, {Mardones}, {Myers}, {Caselli},
  {Bachiller}  \& {Benson}}{{Tafalla} et~al.}{1998}]{1998ApJ...504..900T}
{Tafalla} M.,  {Mardones} D.,  {Myers} P.~C.,  {Caselli} P.,  {Bachiller} R.,
  {Benson} P.~J.,  1998, \mn@doi [\apj] {10.1086/306115}, \href
  {https://ui.adsabs.harvard.edu/abs/1998ApJ...504..900T} {504, 900}

\bibitem[\protect\citeauthoryear{{Takakuwa}, {Tsukamoto}, {Saigo}  \&
  {Saito}}{{Takakuwa} et~al.}{2018}]{2018ApJ...865...51T}
{Takakuwa} S.,  {Tsukamoto} Y.,  {Saigo} K.,   {Saito} M.,  2018, \mn@doi
  [\apj] {10.3847/1538-4357/aadb93}, \href
  {https://ui.adsabs.harvard.edu/abs/2018ApJ...865...51T} {865, 51}

\bibitem[\protect\citeauthoryear{{Triaud}}{{Triaud}}{2018}]{2018haex.bookE...2T}
{Triaud} A. H.~M.~J.,  2018, {The Rossiter-McLaughlin Effect in Exoplanet
  Research}.
p.~2, \mn@doi{10.1007/978-3-319-55333-7_2}

\bibitem[\protect\citeauthoryear{{Tsukamoto} \& {Machida}}{{Tsukamoto} \&
  {Machida}}{2011}]{2011MNRAS.416..591T}
{Tsukamoto} Y.,  {Machida} M.~N.,  2011, \mn@doi [\mnras]
  {10.1111/j.1365-2966.2011.19081.x}, \href
  {http://ads.nao.ac.jp/abs/2011MNRAS.416..591T} {416, 591}

\bibitem[\protect\citeauthoryear{{Tsukamoto} \& {Machida}}{{Tsukamoto} \&
  {Machida}}{2013}]{2013MNRAS.428.1321T}
{Tsukamoto} Y.,  {Machida} M.~N.,  2013, \mn@doi [\mnras]
  {10.1093/mnras/sts111}, \href {http://ads.nao.ac.jp/abs/2013MNRAS.428.1321T}
  {428, 1321}

\bibitem[\protect\citeauthoryear{{Tsukamoto}, {Machida}  \&
  {Inutsuka}}{{Tsukamoto} et~al.}{2013}]{2013MNRAS.436.1667T}
{Tsukamoto} Y.,  {Machida} M.~N.,   {Inutsuka} S.,  2013, \mn@doi [\mnras]
  {10.1093/mnras/stt1684}, \href
  {http://adsabs.harvard.edu/abs/2013MNRAS.436.1667T} {436, 1667}

\bibitem[\protect\citeauthoryear{{Tsukamoto}, {Takahashi}, {Machida}  \&
  {Inutsuka}}{{Tsukamoto} et~al.}{2015a}]{2015MNRAS.446.1175T}
{Tsukamoto} Y.,  {Takahashi} S.~Z.,  {Machida} M.~N.,   {Inutsuka} S.,  2015a,
  \mn@doi [\mnras] {10.1093/mnras/stu2160}, \href
  {http://adsabs.harvard.edu/abs/2015MNRAS.446.1175T} {446, 1175}

\bibitem[\protect\citeauthoryear{{Tsukamoto}, {Iwasaki}, {Okuzumi}, {Machida}
  \& {Inutsuka}}{{Tsukamoto} et~al.}{2015b}]{2015MNRAS.452..278T}
{Tsukamoto} Y.,  {Iwasaki} K.,  {Okuzumi} S.,  {Machida} M.~N.,   {Inutsuka}
  S.,  2015b, \mn@doi [\mnras] {10.1093/mnras/stv1290}, \href
  {http://ads.nao.ac.jp/abs/2015MNRAS.452..278T} {452, 278}

\bibitem[\protect\citeauthoryear{{Tsukamoto}, {Iwasaki}, {Okuzumi}, {Machida}
  \& {Inutsuka}}{{Tsukamoto} et~al.}{2015c}]{2015ApJ...810L..26T}
{Tsukamoto} Y.,  {Iwasaki} K.,  {Okuzumi} S.,  {Machida} M.~N.,   {Inutsuka}
  S.,  2015c, \mn@doi [\apjl] {10.1088/2041-8205/810/2/L26}, \href
  {http://ads.nao.ac.jp/abs/2015ApJ...810L..26T} {810, L26}

\bibitem[\protect\citeauthoryear{{Tsukamoto}, {Okuzumi}, {Iwasaki}, {Machida}
  \& {Inutsuka}}{{Tsukamoto} et~al.}{2017}]{2017PASJ...69...95T}
{Tsukamoto} Y.,  {Okuzumi} S.,  {Iwasaki} K.,  {Machida} M.~N.,   {Inutsuka}
  S.,  2017, \mn@doi [Publications of the Astronomical Society of Japan]
  {10.1093/pasj/psx113}, \href
  {https://ui.adsabs.harvard.edu/abs/2017PASJ...69...95T} {69, 95}

\bibitem[\protect\citeauthoryear{{Tsukamoto}, {Okuzumi}, {Iwasaki}, {Machida}
  \& {Inutsuka}}{{Tsukamoto} et~al.}{2018}]{2018ApJ...868...22T}
{Tsukamoto} Y.,  {Okuzumi} S.,  {Iwasaki} K.,  {Machida} M.~N.,   {Inutsuka}
  S.,  2018, \mn@doi [\apj] {10.3847/1538-4357/aae4dc}, \href
  {https://ui.adsabs.harvard.edu/abs/2018ApJ...868...22T} {868, 22}

\bibitem[\protect\citeauthoryear{{Ward-Thompson}, {Andr{\'e}}, {Crutcher},
  {Johnstone}, {Onishi}  \& {Wilson}}{{Ward-Thompson}
  et~al.}{2007}]{2007prpl.conf...33W}
{Ward-Thompson} D.,  {Andr{\'e}} P.,  {Crutcher} R.,  {Johnstone} D.,  {Onishi}
  T.,   {Wilson} C.,  2007, in {Reipurth} B.,  {Jewitt} D.,   {Keil} K.,  eds,
  Protostars and Planets V. p.~33 (\mn@eprint {arXiv} {astro-ph/0603474})

\bibitem[\protect\citeauthoryear{{Williams}, {Myers}, {Wilner}  \& {Di
  Francesco}}{{Williams} et~al.}{1999}]{1999ApJ...513L..61W}
{Williams} J.~P.,  {Myers} P.~C.,  {Wilner} D.~J.,   {Di Francesco} J.,  1999,
  \mn@doi [\apjl] {10.1086/311895}, \href
  {https://ui.adsabs.harvard.edu/abs/1999ApJ...513L..61W} {513, L61}

\bibitem[\protect\citeauthoryear{{Winn} \& {Fabrycky}}{{Winn} \&
  {Fabrycky}}{2015}]{2015ARA&A..53..409W}
{Winn} J.~N.,  {Fabrycky} D.~C.,  2015, \mn@doi [\araa]
  {10.1146/annurev-astro-082214-122246}, \href
  {https://ui.adsabs.harvard.edu/abs/2015ARA&A..53..409W} {53, 409}

\bibitem[\protect\citeauthoryear{{Winn} et~al.,}{{Winn}
  et~al.}{2005}]{2005ApJ...631.1215W}
{Winn} J.~N.,  et~al., 2005, \mn@doi [\apj] {10.1086/432571}, \href
  {https://ui.adsabs.harvard.edu/\#abs/2005ApJ...631.1215W} {631, 1215}

\bibitem[\protect\citeauthoryear{{Wurster} \& {Bate}}{{Wurster} \&
  {Bate}}{2019}]{2019MNRAS.tmp..933W}
{Wurster} J.,  {Bate} M.~R.,  2019, \mn@doi [\mnras] {10.1093/mnras/stz1023},
  \href {https://ui.adsabs.harvard.edu/abs/2019MNRAS.tmp..933W} {p.~933}

\bibitem[\protect\citeauthoryear{{Wurster}, {Price}  \& {Bate}}{{Wurster}
  et~al.}{2016}]{2016MNRAS.457.1037W}
{Wurster} J.,  {Price} D.~J.,   {Bate} M.~R.,  2016, \mn@doi [\mnras]
  {10.1093/mnras/stw013}, \href {http://ads.nao.ac.jp/abs/2016MNRAS.457.1037W}
  {457, 1037}

\bibitem[\protect\citeauthoryear{{Wurster}, {Price}  \& {Bate}}{{Wurster}
  et~al.}{2017}]{2017MNRAS.466.1788W}
{Wurster} J.,  {Price} D.~J.,   {Bate} M.~R.,  2017, \mn@doi [\mnras]
  {10.1093/mnras/stw3181}, \href
  {http://adsabs.harvard.edu/abs/2017MNRAS.466.1788W} {466, 1788}

\bibitem[\protect\citeauthoryear{{Wurster}, {Bate}  \& {Price}}{{Wurster}
  et~al.}{2018}]{2018MNRAS.475.1859W}
{Wurster} J.,  {Bate} M.~R.,   {Price} D.~J.,  2018, \mn@doi [\mnras]
  {10.1093/mnras/stx3339}, \href {http://ads.nao.ac.jp/abs/2018MNRAS.475.1859W}
  {475, 1859}

\bibitem[\protect\citeauthoryear{{Xue} \& {Suto}}{{Xue} \&
  {Suto}}{2016}]{2016ApJ...820...55X}
{Xue} Y.,  {Suto} Y.,  2016, \mn@doi [\apj] {10.3847/0004-637X/820/1/55}, \href
  {https://ui.adsabs.harvard.edu/\#abs/2016ApJ...820...55X} {820, 55}

\bibitem[\protect\citeauthoryear{{Xue}, {Suto}, {Taruya}, {Hirano}, {Fujii}  \&
  {Masuda}}{{Xue} et~al.}{2014}]{2014ApJ...784...66X}
{Xue} Y.,  {Suto} Y.,  {Taruya} A.,  {Hirano} T.,  {Fujii} Y.,   {Masuda} K.,
  2014, \mn@doi [\apj] {10.1088/0004-637X/784/1/66}, \href
  {https://ui.adsabs.harvard.edu/\#abs/2014ApJ...784...66X} {784, 66}

\bibitem[\protect\citeauthoryear{{Yamamoto}}{{Yamamoto}}{2017}]{2017iace.book.....Y}
{Yamamoto} S.,  2017, {Introduction to Astrochemistry: Chemical Evolution from
  Interstellar Clouds to Star and Planet Formation},
  \mn@doi{10.1007/978-4-431-54171-4.
}

\bibitem[\protect\citeauthoryear{{Yen} et~al.,}{{Yen}
  et~al.}{2014}]{2014ApJ...793....1Y}
{Yen} H.-W.,  et~al., 2014, \mn@doi [\apj] {10.1088/0004-637X/793/1/1}, \href
  {https://ui.adsabs.harvard.edu/abs/2014ApJ...793....1Y} {793, 1}

\bibitem[\protect\citeauthoryear{{Yen}, {Koch}, {Takakuwa}, {Krasnopolsky},
  {Ohashi}  \& {Aso}}{{Yen} et~al.}{2017}]{2017ApJ...834..178Y}
{Yen} H.-W.,  {Koch} P.~M.,  {Takakuwa} S.,  {Krasnopolsky} R.,  {Ohashi} N.,
  {Aso} Y.,  2017, \mn@doi [\apj] {10.3847/1538-4357/834/2/178}, \href
  {https://ui.adsabs.harvard.edu/abs/2017ApJ...834..178Y} {834, 178}

\bibitem[\protect\citeauthoryear{{Yoneda}, {Tsukamoto}, {Furuya}  \&
  {Aikawa}}{{Yoneda} et~al.}{2016}]{2016ApJ...833..105Y}
{Yoneda} H.,  {Tsukamoto} Y.,  {Furuya} K.,   {Aikawa} Y.,  2016, \mn@doi
  [\apj] {10.3847/1538-4357/833/1/105}, \href
  {http://adsabs.harvard.edu/abs/2016ApJ...833..105Y} {833, 105}

\makeatother
\end{thebibliography}
\input{takaishids_SPH1_for_arxiv_20200116.bbl}

%%%%%%%%%%%%%%%%%%%%%%%%%%%%%%%%%%%%%%%%%%%%%%%%%%

%%%%%%%%%%%%%%%%% APPENDICES %%%%%%%%%%%%%%%%%%%%%

\appendix

\section{Numerical convergence test with respect to
 the Different Mass, Realization of Turbulence, and Number of SPH
 particles}
\label{app:NumericalConvergence}
%%%%%%%%%%%%%%%%%%%%%%%%%%%%%%

In this appendix, we discuss whether the factors which are not
considered in this paper change our conclusion or not.  For this
purpose, we performed the simulations with (1) the different
realization of the turbulence, (2) the different mass ($0.3\msun,
3\msun$), and (3) the different number of the SPH particles of the
initial cloud core for our fiducial model \FiducialModel.

\begin{figure}%%% Figure App1
  \centering
  \includegraphics[clip,width=60mm,angle=-90]{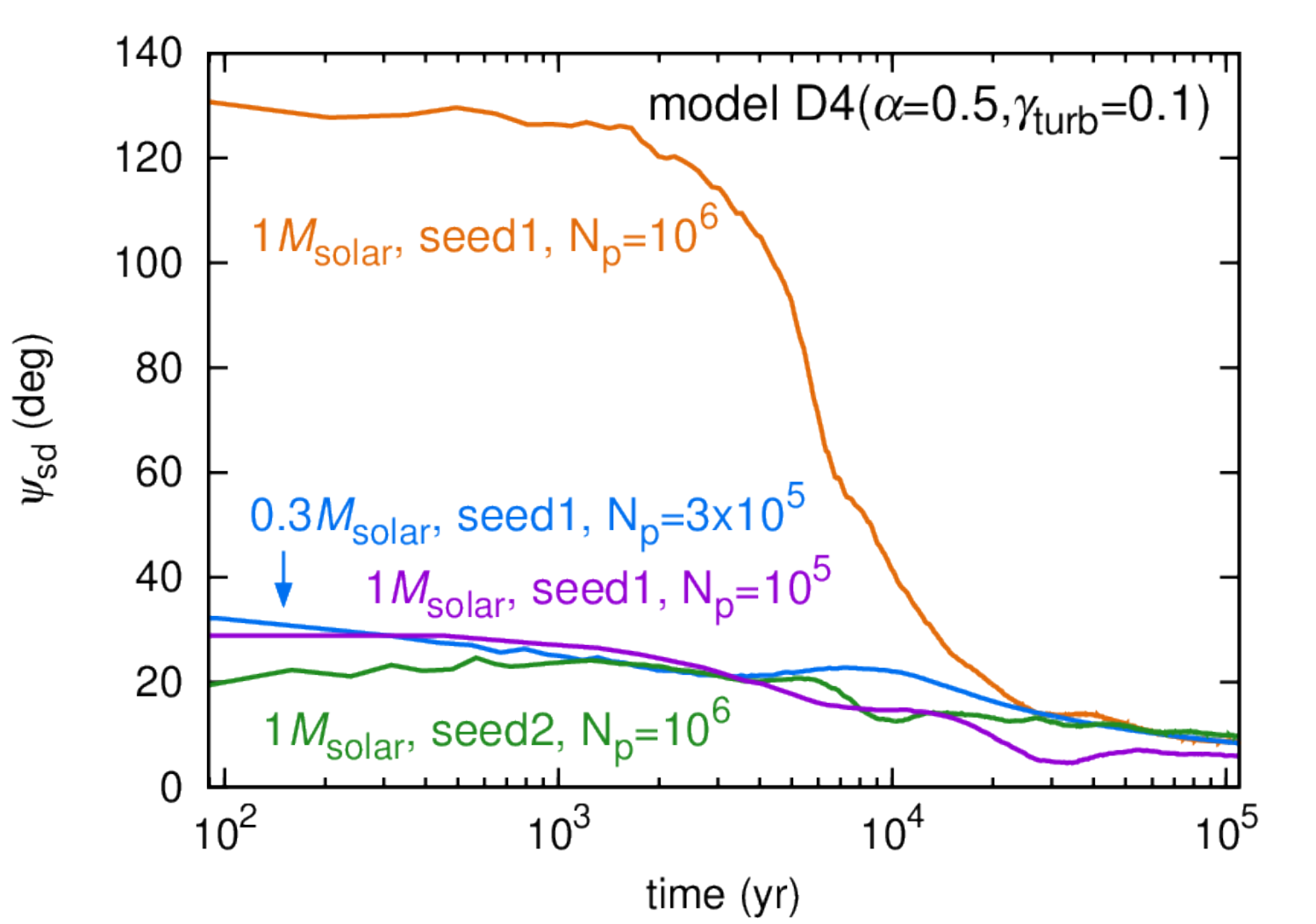}
  \caption{ Time evolution of $\psd$ for our fiducial model
    \FiducialModel in which the mass, realization of the turbulence,
    and number of SPH particles of the initial cloud core are
    different.
    \label{fig:app1_angle_diff_mass_and_seed}
  }
\end{figure}

Figure \ref{fig:app1_angle_diff_mass_and_seed} summarize our results.
At first, we check the impact of the different realization of the
turbulence of the initial cloud core.  In this study, we only consider
one realization of the turbulence for one parameter set of $\alpha$
and $\gturb$.  However, due to its stochastic nature, the different
realization may causes the different conclusions.  The green solid
line in Figure \ref{fig:app1_angle_diff_mass_and_seed} plots the time
evolution of $\psd$ with the same parameters of our fiducial model
\FiducialModel but varying the realization of the turbulence (model
Seed2) and shows that the initial star-disk angle $\psd$ is much
smaller than that of model D4.  This is not surprising because the
different realization of the turbulence changes the initial
distribution of the angular momentum around the protostar.  Thus the
stellar spin direction at its formation epoch is significantly
affected by the realization.  Note however, that $\psd$ converged to
less than $20^{\circ}$ due to the mechanism discussed in section
\ref{sec:D4-result}, and our main conclusion is not changed by the
random nature of the turbulence.

Next, we check the impact of the mass of the initial cloud core.  We
conducted here two simulations with the parameters of our fiducial
model \FiducialModel but varying the mass of the initial cloud core as
$0.3\msun$ (model Small) and $3\msun$ (model Large).

One protostar is formed in model Small, and a binary system is formed
in model Large.  Blue solid line in Figure
\ref{fig:app1_angle_diff_mass_and_seed} shows the time evolution of
$\psd$ of model Small in which the mass of the initial cloud core is
$0.3\msun$ and shows that while the initial star-disk angle $\psd$ of
the different mass of $0.3\msun$ is smaller than that of model D4, it
also converged to less than $20^{\circ}$, and there is no significant
misalignment of the star-disk angle $\psd$ of model Small at $t \sim
10^5$ yr.  Therefore our main conclusion is still maintained with this
calculation.

A wide binary system with separation $\sim100$ au is formed in model
Large.  Because we will focus on the isolated systems in this work, we
do not discuss the result of model Large here.  Note, however, that
the multiplicity strongly depends on the mass of the cloud core even
with the same parameter of $\alpha$ and $\gturb$. 

Finally, we check the impact of the numerical resolution.  For this
purpose, we conducted a simulation with the parameters of our fiducial
model \FiducialModel but varying the number of the SPH particles of
the initial cloud core as $N_{\rm p} \sim 10^5$ (model Low).

Violet solid line in Figure \ref{fig:app1_angle_diff_mass_and_seed}
plots the time evolution of $\psd$ of model Low and shows that the
initial star-disk angle $\psd$ of model Low is also smaller than that
of model D4 likewise the case of the model Seed2, and it also
converged to less than $20^{\circ}$.  It may be related to the change
of the interpolation of the initial turbulent velocity field caused by
the different initial resolution.  The star-disk angle $\psd$ also
converges to less than $20^{\circ}$ in $\sim 10^{4}$ yr after the
protostar formation.  Therefore, our main conclusion is still
maintained with the smaller numerical resolution.

%%%%%%%%%%%%%%%%%%%%%%%%%%%%%%
%%%%%%%%%%%%%%%%%%%%%%%%%%%%%%

%%%%%%%%%%%%%%%%%%%%%%%%%%%%%%
\begin{table*}%%% Table a1
  \centering
  \caption{ 
    Initial parameters of the molecular cloud cores
    for the calculations in Appendix A;
    $\beff$ is the dimensionless angular momentum,
    $R_{\rm init}$, $M_{\rm init}$, 
    and $\rho_{\rm init}=3M_{\rm init}/(4\pi R_{\rm init}^3)$ are 
    the initial radius, mass, and density of the initial cloud cores.
    Seed (a random number) is used for
    the implementation of the initial turbulent velocity field.
    ${\cal M}$ is the initial Mach number,
    $t_{\rm ff}=\sqrt{3\pi/(32G\rho_{\rm init})}$ is the free-fall time and
    $N_{\rm p}$ is the number of SPH particles of the initial cloud core.
    $\psd(t=10^2\rm{yr})$ and $\psd(t=10^5\rm{yr})$ which are the
    star-disk angles at $t=10^2$ yr and $t=10^5$ yr are listed in
    the third from the end and penultimate columns, respectively.
    The last column indicates the multiplicity of the protostars in our simulations. 
  }
  \scalebox{0.9}[1.0]{ % size down {row- size}[column| size]
    \begin{tabular}{cccccccccccc}
      \hline
      \hline
      Model & $\beff$ & $R_{\rm init}~[\rm au]$ & $M_{\rm init}$ & $\rho_{\rm init}~[{\rm g~cm^{-3}}]$ &
      seed & $\mach$ & $t_{\rm ff}~[\rm yr]$ & $N_{\rm p}$ &
      $\psd(t=10^2\rm{yr})$ & $\psd(t=10^5\rm{yr})$ & multiplicity \\
      \hline
      D4    & 0.012  & 4917  & $1\msun$   & $1.2\times10^{-18}$ & seed1 & 0.77 &
      $6.1\times10^{4}$ & $1,045,414\approx 10^6$ &
      $127.7^{\circ}$ & $8.7^{\circ}$ & single \\
      Seed2 & 0.012  & 4917  & $1\msun$   & $1.2\times10^{-18}$ & seed2 & 0.77 &
      $6.1\times10^{4}$ & $1,045,414\approx 10^6$ &
    $22.3^{\circ}$  & $9.9^{\circ}$ & single \\
      Low   & 0.012  & 4947  & $1\msun$   & $1.2\times10^{-18}$ & seed1 & 0.77 &
      $6.1\times10^{4}$ & $~104,470~\approx 10^5$ &
      $28.9^{\circ}$  & $6.1^{\circ}$ & single \\
      Small & 0.0011 & 1475  & $0.3\msun$ & $1.3\times10^{-17}$ & seed1 & 0.77 &
      $1.8\times10^{4}$ & $~313,858~\approx 3\times10^5$ &
      $30.1^{\circ}$  & $8.6^{\circ}$ & single \\
      Large & 0.059  & 14750 & $3\msun$   & $1.3\times10^{-19}$ & seed1 & 0.77 &
      $1.8\times10^{5}$ & $3,140,355\approx 3\times10^6$ &
      & - & binary \\
      \hline
      \hline
    \end{tabular}
  } %  \scalebox{0.9}[1.0]{ % size down {row- size}[column| size]
  \label{tb:app1_initial_conditions}
  \footnotesize
\end{table*}
%%%%%%%%%%%%%%%%%%%%%%%%%%%%%%

%%%%%%%%%%%%%%%%%%%%%%%%%%%%%%

%%%%%%%%%%%%%%%%%%%%%%%%%%%%%%%%%%%%%%%%%%%%%%%%%%

% Don't change these lines
\bsp	% typesetting comment
\label{lastpage}
\end{document}